\documentclass[aps,prd,amsmath,amssymb,10pt,letterpaper,balancelastpage,showpacs,superscriptaddress,nofootinbib,floatfix,notitlepage,longbibliography,twocolumn
]{revtex4-2}
\usepackage{dsfont}
\usepackage{graphicx}
\usepackage{dcolumn}
\usepackage{bm}
\usepackage{color}
\usepackage{epsfig}
\usepackage{hyperref}
\usepackage{natbib}
\usepackage{float}
\usepackage{footmisc}
\usepackage{slashed}
\usepackage{comment}
\usepackage{aas_macros}
\definecolor{lightred}{rgb}{1,0.5,0.5}
\definecolor{lightgreen}{rgb}{0.5,1,0.5}
\definecolor{lightblue}{rgb}{0.5,0.5,1}
\definecolor{lightcyan}{rgb}{0.5,0.75,0.75}
\definecolor{lightmagenta}{rgb}{0.75,0.5,0.75}
\definecolor{customgreen}{rgb}{0.494,1,0.502}

\newcommand{\meV}{\mathinner{\mathrm{meV}}}
\newcommand{\eV}{\mathinner{\mathrm{eV}}}
\newcommand{\keV}{\mathinner{\mathrm{keV}}}
\newcommand{\MeV}{\mathinner{\mathrm{MeV}}}
\newcommand{\GeV}{\mathinner{\mathrm{GeV}}}

\begin{document}

\title{Cosmological Limits on Strong Dark Forces}

\author{Peter W.~Graham}
\affiliation{Leinweber Institute for Theoretical Physics, Department of Physics, Stanford University, Stanford, CA 94305, USA}
\affiliation{Kavli Institute for Particle Astrophysics and Cosmology, Department of Physics, Stanford University, Stanford, CA 94305, USA}

\author{Harikrishnan Ramani}
\affiliation{Department of Physics and Astronomy, University of Delaware and the Bartol Research Institute, Newark, DE 19716, USA}

\author{Olivier Simon}
\affiliation{Princeton Center for Theoretical Science, Princeton University, Princeton, NJ 08544, U.S.A.}
\affiliation{Department of Physics, Princeton University, Princeton, NJ 08544, U.S.A.}

\author{Erwin H.~Tanin}
\email{ehtanin@stanford.edu}
\affiliation{Leinweber Institute for Theoretical Physics, Department of Physics, Stanford University, Stanford, CA 94305, USA}

\begin{abstract}
We showcase cosmology's ability to constrain long-range forces between dark matter particles. Specifically, we consider a fermionic dark matter interacting via a Yukawa-coupled light scalar, focusing on regimes where the dark forces are stronger than gravitational and yet unconstrained. We show that the dark sector dynamics, both at the background and perturbation levels, is far richer than what can be captured with just the static interparticle Yukawa potential. The background dynamics includes an attractor that funnels a wide range of initial conditions onto an evolution unique to each parameter space. In a large swath of parameter space beyond existing limits, the dark sector deviates drastically from cold dark matter in observable epochs. We rule out this parameter space using existing constraints on dark-sector equation of state and small-scale cosmic perturbations, thus setting the strongest constraints yet on dark matter self-interactions at length scales shorter than 100 kpc. In addition, we briefly discuss repulsive dark forces and place cosmological limits that are stricter than in the attractive case. 
\end{abstract}

\maketitle
\tableofcontents

\clearpage
\section{Introduction}

Decades of dark matter (DM) searches have placed increasingly stringent bounds on its possible interactions with the Standard Model (SM). We are gradually approaching the sobering possibility that any non-gravitational DM - SM coupling, if it exists, is too feeble to be detected by current or near-future experiments. Even in this ``nightmare'' scenario, the dark sector may still exhibit rich internal dynamics. Such dynamics would manifest as deviations from the $\Lambda$CDM paradigm, most notably through imprints on small-scale structure or via deviations from the well measured DM equation of state. Examples of things learned about the dark sector from such probes include lower limits on  the dark matter particle mass~\cite{Rogers:2020ltq}, the earliest epoch by which dark matter needed to be cold~\cite{Das:2020nwc}, dark matter lifetime~\cite{poulin2016fresh}, stringent limits on its primordial power spectrum~\cite{Graham:2024hah}, post inflationary PQ breaking scenarios~\cite{Chang:2024fol}, the possible presence of strong self-interactions at small scales~\cite{Tulin:2017ara} as well as tight constraints~\cite{Slone:2021nqd} at these scales to name a few.

A particularly intriguing  non-triviality is that dark matter experiences a long-range ``fifth force.'' These forces are typically mediated by an ultralight scalar field, motivated in scenarios such as models with extra dimensions~\cite{Sundrum:2003yt} or as moduli arising naturally in string theory~\cite{damour1994string, damour1994string1}. Fifth forces between SM particles have been extensively searched for, with experimental constraints requiring them to be weaker than gravity down to the $\mathcal{O}(10~\mu\mathrm{m})$ scale. In general, such forces can violate the equivalence principle, and in the extreme case, may be confined entirely within the dark sector rendering them invisible to conventional probes.

Astrophysical considerations have constrained the strength of long-range forces between DM particles (dark forces) with ranges longer than about $1\text{ Mpc}$ to be weaker than gravity \cite{Bogorad:2023wzn}; see also \cite{Keselman:2009oaz,Kesden:2006zb,Kesden:2006vz,Frieman:1991zxc,1992ApJ...398..407G}. However in the sub-Mpc range, 
dark forces with strengths orders of magnitude stronger than gravity are thus far still allowed~\cite{Bogorad:2023wzn}. Such dark forces could make drastic changes to the early universe, when it was much smaller and denser. On the other hand, consistency with a wealth of cosmological data necessitates that the DM behaves like cold dark matter (CDM) beginning at redshifts of $z\sim 10^5$, whereupon the universe was about 100 pc in size. Thus, the early universe could in principle probe dark forces as short as 100 pc in range at near-gravitational strengths.

There have been efforts to constrain long-range forces between dark matter particles through their cosmological consequences \cite{Farrar:2003uw,Bean:2008ac, Bean:2007ny,2010PhRvD..81f3521K,Hellwing:2008qf,Bottaro:2024pcb,Bottaro:2023wkd,Archidiacono:2022iuu,Costa:2025kwt}. In early studies \cite{Farrar:2003uw,Bean:2008ac, Bean:2007ny,2010PhRvD..81f3521K,Hellwing:2008qf}, the interaction between two DM particles with mass $m_\chi$ separated by a distance $r$ is simply modeled as a static Yukawa potential
\begin{align}
    |V_{\rm Yuk}|=\frac{\alpha_{\chi\chi}Gm_\chi^2 }{r}e^{-r/\lambda_\phi},\label{eq:YukForce}
\end{align}
where $\lambda_\phi$ and $\alpha_{\chi\chi}$ parametrize, respectively, the range of the dark force and its strength relative to gravity. A particle-physics realization of such a force requires a light field serving as a mediator. The potential in Eq.~\eqref{eq:YukForce} corresponds to the static solution of the mediator field, analogous to electrostatics. While this approximation may be appropriate in static systems such as those of small-scale tests of gravity, it may not adequately describe the mediator field in the early universe, which typically has its own rich dynamics that is not completely dictated by the DM configuration at an instance \cite{Domenech:2021uyx,Domenech:2023afs}. 

More recently, refs.~\cite{Bottaro:2024pcb,Bottaro:2023wkd,Archidiacono:2022iuu,Costa:2025kwt} carefully considered the dynamical nature of the mediator field in placing limits on dark forces. Their findings ruled out dark forces with
ranges $\lambda_\phi\gtrsim 100\text{ kpc}$ and strengths relative to gravity
$\alpha_{\chi\chi}\gtrsim 10^{-3}-10^{-2}$, thereby placing the most stringent constraint in this regime of the parameter space. Note, however, that these analyses are based on a perturbative expansion in the coupling $\alpha_{\chi\chi}$ and are therefore restricted, for forces in those ranges, to dark forces with strengths at most comparable to that of gravity, $\alpha_{\chi\chi}\lesssim 1$.

Our aim in this paper is to determine cosmological constraints on long-range dark forces with coupling strengths larger than gravitational, beyond the current reach of astrophysical observations and small-coupling analyses of cosmology. We will thus focus mainly on new forces with range $\lambda_\phi\lesssim 100\text{ kpc}$ and $\alpha_{\chi\chi}\gtrsim 1$. While a detailed likelihood analysis is crucial to disentangle the effects of dark forces with small-coupling  from that of other cosmological parameters, large-coupling dark forces may lead to drastic modifications to the standard cosmology well beyond what is allowed by the inherent degeneracies of cosmological parameters. We focus on the following striking effects of dark forces. First, the interplay between the DM and the mediator field generically leads to a dynamical dark sector equation of state that, at times, deviates far from that of CDM. Second, long-range dark forces could cause orders of magnitude of extra growth in observable cosmological perturbation modes. The latter is an effect known to occur when the quasi-static exchange interaction between dark matter particles, Eq.~\eqref{eq:YukForce}, dominates dark sector dynamics \cite{Afshordi:2005ym,Domenech:2023afs,Savastano:2019zpr}, but, as we will discuss, this condition is often not realized in the early universe. Even so, in some cases, a catastrophic amount of growth can take place during the limited period of applicability of Eq.~\eqref{eq:YukForce}. We leverage these phenomena to place conservative cosmological limits on dark forces.

The paper is organized as follows. In Section~\ref{s:assumptions} we describe the Yukawa dark-force model and specify the assumptions that enter our analyses. In Section~\ref{s:background} we discuss the impact of dark forces on background cosmology and how it could be probed. In Section~\ref{s:perturbations} we consider the enhanced growth of cosmic perturbations in the presence of dark forces and how it could be probed. In Section~\ref{s:results} we detail our main results. In Section~\ref{s:repulsive} we briefly discuss repulsive dark forces. Finally, we conclude in Section.~\ref{s:conclusion}.

\clearpage

\section{Assumptions}
\label{s:assumptions}
\subsection{Model}
The Yukawa parametrization of dark forces, Eq.~\eqref{eq:YukForce}, enables a model-independent treatment of static dark forces. However, certain considerations, such as the early-universe dynamics of the mediator field, are inherently model-dependent. For this work, we adopt one of the simplest models of long-range forces, namely a fermion $\chi$ of mass $m_\chi$ serving as the primary DM with a Yukawa coupling $g$ to a light scalar mediator $\phi$ of mass $m_\phi$, as described by the following Lagrangian \cite{Farrar:2003uw}
\begin{align}
    \mathcal{L}_{\rm dark}= \frac{1}{2}\left(\partial \phi\right)^2+\bar{\chi}i\slashed{\partial}\chi-\left(m_\chi-g\phi\right)\bar{\chi}\chi-\frac{1}{2}m_\phi^2\phi^2.\nonumber
\end{align}
In the main part of our analysis, the dark coupling $g$ and the fermion bare mass $m_\chi$ enter only through the combination $g/m_\chi$. As long as the mass $m_\chi$ is limited to certain $(g,m_\phi)$-dependent range, its exact value is not important for our main analysis; see Appendix~\ref{appendix:finitedensity}. Thus, it is convenient to express $g$ in terms of the strength of the dark long-range force relative to that of gravity
\begin{align}
    \alpha_{\chi\chi}\equiv \frac{g^2}{4\pi Gm_\chi^2},
    \label{eq:alpha}
\end{align}
as in Eq.~\ref{eq:YukForce}, where in this model the force range is given by the inverse mass of the mediator $\lambda_\phi= m_\phi^{-1}$. Throughout the paper, we will assume that the dark sector is completely decoupled from the Standard Model sector.

Note that for a \emph{scalar} force mediator, the two-body static exchange force is universally \emph{attractive} between particles and anti-particles of the same kind. This is in contrast to static forces mediated by a vector, such as the electrostatic force in the SM, which are repulsive between particle--particle and anti-particle--anti-particle pairs, but attractive between particle--anti-particle pairs. We briefly consider repulsive dark forces in Section.~\ref{s:repulsive}.

Throughout our main analysis, we neglect the scalar field's quartic coupling, requiring it be smaller than a certain threshold (see Appendix.~\ref{appendix:finitedensity}) such that it can be safely ignored. We note that both the bare scalar masses $m_\phi$ and self-interactions considered in this paper are fine-tuned in the sense that their values are smaller than the expected loop-contributions to them. See Appendix~\ref{Appendix:Naturalness} for more details. However, there are theoretical realizations~\cite{Blinov:2016kte, Hook:2018jle,Brzeminski:2020uhm} that ameliorate such fine-tuning and make the scalar generically light.

\subsection{Initial Condition}
We begin our analysis at some scale factor $a_i$ deep in radiation domination (RD), e.g., at $a_i=10^{-9}$, early enough that the exact value of $a_i$ is unimportant. In the simplest case, we start the $\phi$ background field at the value $\phi_i=0$ and populate $\chi$ completely asymmetrically, with only particles and no antiparticles. We also assume that the quanta of the $\phi$ and $\chi$ fields are sufficiently cold that their temperatures do not affect our analysis. We consider the effects of finite $\chi$ density and finite $\chi$ temperature on the background dynamics of $\phi$ in Appendix~\ref{appendix:finitedensity}. We find that avoiding the regimes where these effects are important amounts to mild restrictions on the fermion mass  $m_\chi$.

Nevertheless, thanks to both the existence of a dynamical attractor in the $\phi$ field's background evolution and the insensitivity of the backreaction of a non-relativistic $\chi$ to its precise momentum distribution, a broad range of initial conditions other than the simplest one, as mentioned in the previous paragraph, converge to virtually the same dynamics at sufficiently late times, relevant for cosmological observations. We describe these further in Appendix~\ref{appendix:scalardetail} and \ref{appendix:finitedensity}. In all the cases we consider, the number density of $\chi$ is covariantly conserved,  $n_\chi\propto a^{-3}$. Thus, we can write its initial value as
\begin{align}
    n_{\chi,i}=\frac{f_\chi\rho_{\rm m,0}}{m_\chi}\left(\frac{a_i}{a_0}\right)^{-3}.\label{eq:fchi}
\end{align}
Here, the subscripts $_i$ and $_0$ refer to the initial and present-day values, and $\rho_{\rm m,0}\approx 12.5\,\meV^4$ is the fiducial cosmic matter density today \cite{Planck:2018vyg}.  For the most part, we will assume that $f_\chi=f_{\rm DM}=0.85$, corresponding to the current DM mass density of $\rho_{\rm DM,0}=f_{\rm DM}\rho_{\rm m,0}=10.6\,\meV^4$. An exception to this occurs in regimes that predict the $\phi$ field behaves like an ultralight CDM component and contributes significantly to the DM abundance at the present epoch, in which case the $f_\chi$ should be appropriately calibrated such that the total mass density of $\phi$ and $\chi$ (not just $\chi$'s) equals $\rho_{\rm DM,0}$ at the present epoch; we will return to this subtlety in Section.~\ref{ss:mixedDM}.

\section{Background Cosmology}
\label{s:background}

\subsection{Effective Potential}

\begin{figure}
    \centering
    \includegraphics[width=\linewidth]{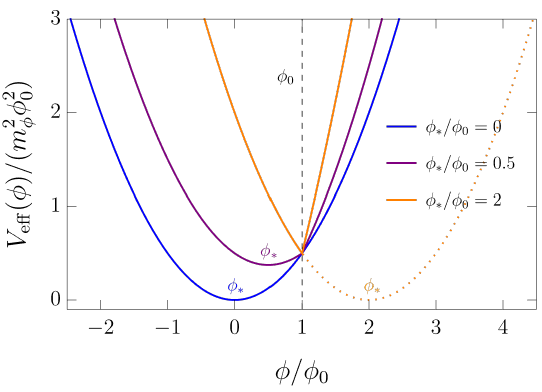}
    \caption{Effective potential of $\phi$, c.f.~Eq.~\eqref{eq:Veff}. The labels ``$\phi_0$" and ``$\phi_*$" mark the locations of the linear finite-density minimum $\phi_0=m_\chi/g$ and the quadratic finite-density minimum $\phi_*=gn_\chi/m_\phi^2$. In the $\phi_*/\phi_0=2$ case, the $\phi_*$ exists mathematically (the dotted line shows what $V_{\rm eff}$ would be if the $|\phi_0-\phi|$ in Eq.~\eqref{eq:Veff} is replaced with $\phi_0-\phi$ without the absolute value) but in this case $\phi_*$ cannot be reached, because before $\phi$ can get there, the slope of the linear potential flips sign at $\phi=\phi_0$. }
    \label{fig:Veff}
\end{figure}

\begin{figure*}[t!]
    \centering
    \includegraphics[width=0.85\linewidth]{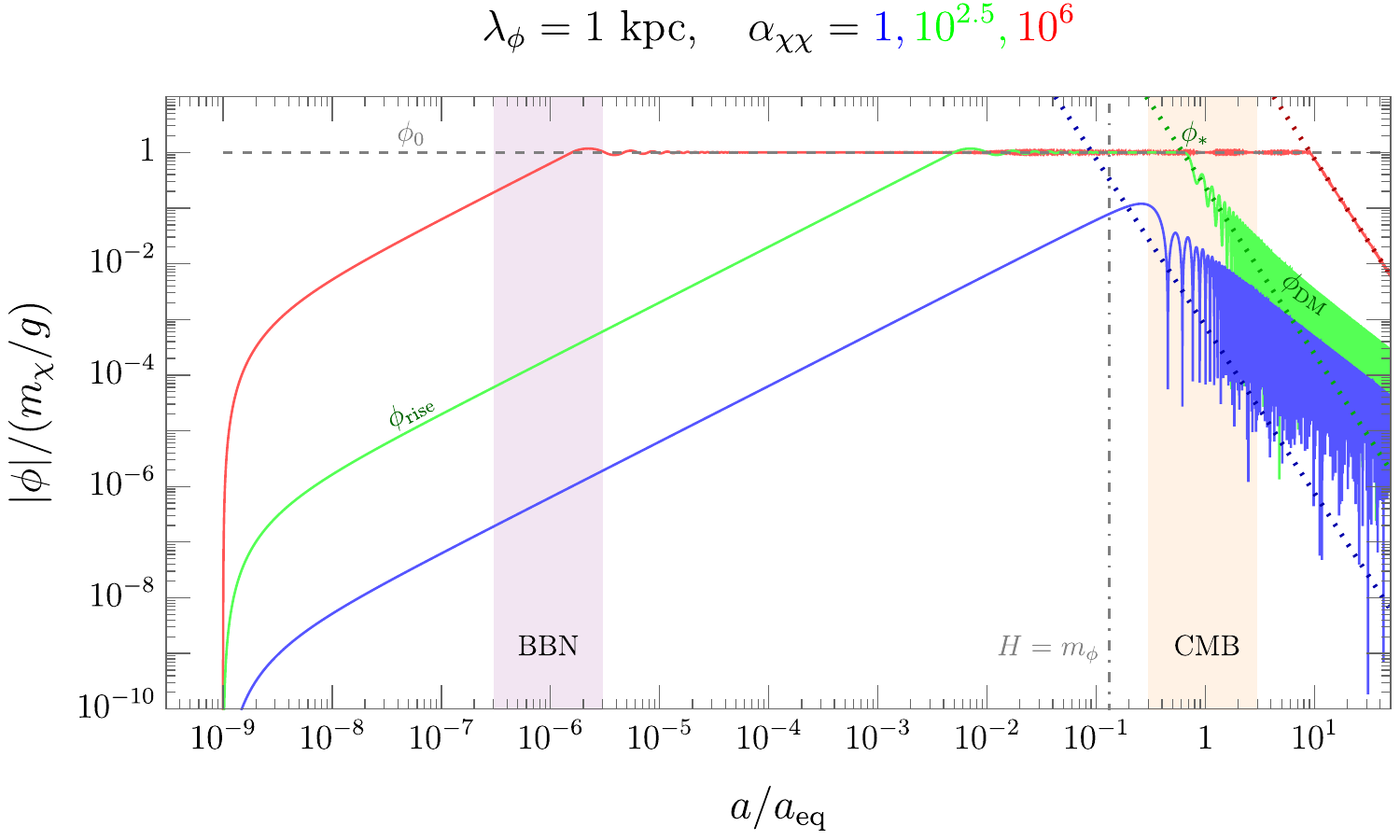}
    \caption{Evolution of the background scalar field $\phi$ as a function of the scale factor $a$ for $\lambda_\phi=1\text{ kpc}$ and $\alpha_{\chi\chi}=1$ (solid blue), $\alpha_{\chi\chi}=10^{2.5}$ (solid green), $\alpha_{\chi\chi}=10^6$ (solid red). Here, we set $a_i=10^{-9}$, $\phi(a_i)=0$, and $\dot{\phi}(a_i)=0$, and numerically evolve $\phi$ using the equation of motion Eq.~\ref{eq:EoM}. The gray dashed line is the linear finite-density minimum $\phi_0\equiv m_\chi/g$ defined in Eq.~\eqref{eq:phi0}. The dotted lines are the quadratic finite-density minima $\phi_*\equiv gn_\chi/m_\phi^2$, defined in Eq.~\eqref{eq:phistar}, corresponding to the values of $\alpha_{\chi\chi}$ matching their colors. The dark green labels ``$\phi_{\rm rise}$" and ``$\phi_{\rm DM}$" indicate, respectively, where in the evolution of the green line the $\phi$ field tracks the attractor solution $\phi_{\rm rise}\equiv (3\alpha_{\chi\chi}f_\chi/4)\phi_0(a/a_{\rm eq})$, defined in Eq.~\eqref{eq:phirise}, and oscillates as CDM, with an amplitude scaling as $\propto a^{-3/2}$. The colored bands labeled ``BBN" and ``CMB" indicate the very rough points where BBN and recombination take place.}
    \label{fig:phievolution}
\end{figure*}

In the presence of a background $\phi$, the $\chi$ particles receives a mass correction such that its effective mass becomes\footnote{Note that the sign of a fermion mass term is unphysical because it can be flipped by a chiral rotation $\chi\rightarrow e^{i\gamma_5\pi/2}\chi$ which leaves the rest of the Lagrangian invariant. The two theories related by this rotation are identical, so no physical quantity can depend on the sign of $m_\chi-g\phi$, only its magnitude $|m_\chi-g\phi|$ is meaningful. For instance, the dispersion relation of $\chi$ in a $\phi$ background reads $E=\sqrt{p^2+(m_\chi-g\phi)^2}$, which is manifestly independent of the sign of $m_\chi-g\phi$.}
\begin{align}
    M_\chi(\phi)=|m_\chi-g\phi|.
\end{align}
As long as $\chi$ is non-relativistic, we can approximate $(m_\chi-g\phi)\left<\bar{\chi}\chi\right>\approx M_\chi(\phi)n_\chi$.\footnote{Since fermions and bosons behave similarly when they are non-relativistic and non-overlapping in wavefunction, many of our conclusions will also apply e.g.~to a real scalar $\tilde{\chi}$ in place of our fermion DM $\chi$, with an effective mass in the presence of $\phi$ background mimicking that of our fermion $\chi$, namely $M_{\tilde{\chi}}(\phi)=m_{\tilde\chi}-g\phi$. This specific effective mass of the scalar $\tilde{\chi}$ can arise for example from the Lagrangian term  $-\mathcal{L}\supset (m_{\tilde{\chi}}-g\phi)^2\tilde{\chi}^2/2$ \cite{Farrar:2003uw}. } The effective potential of $\phi$ is then a sum of a quadratic piece and a linear piece whose slope changes sign at $\phi=m_\chi/g$
\begin{align}
    V_{\rm eff}(\phi)&=\frac{1}{2}m_\phi^2\phi^2+(m_\chi-g\phi)\left<\bar{\chi}\chi\right>\nonumber\\
    &\approx\frac{1}{2}m_\phi^2\phi^2+m_\phi^2\phi_*|\phi_0-\phi \label{eq:Veff}|.
\end{align}
 Here, we have defined the following finite-density minima
\begin{align}
    \phi_0&\equiv\frac{m_\chi}{g}\quad \text{(linear  minimum)},\label{eq:phi0}\\
    \phi_*&\equiv\frac{gn_\chi}{m_\phi^2}\quad \text{(quadratic minimum)}.\label{eq:phistar}
\end{align} 
In Appendix.~\ref{appendix:finitedensity}, we consider a more general finite-density effective potential of $\phi$ and clarify the regime of validity of the above Eq.~\eqref{eq:Veff}. We plot in Fig.~\ref{fig:Veff} the $V_{\rm eff}(\phi)$ for representative values of $\phi_*/\phi_0$. It can be seen that there are two types of minima, $\phi_*$ and $\phi_0$, which we dub the quadratic minimum and linear minimum, respectively. The linear potential piece $m_\phi^2\phi_*|\phi_0-\phi|$ arises from the tendency of the $\phi$-$\chi$ system to minimize the non-relativistic $\chi$'s effective mass $|m_\chi-g\phi|$ at zero, which occurs at the linear minimum $\phi=\phi_0$.\footnote{The approximation that  $(m_\chi-g\phi)\left<\bar{\chi}\chi\right>$ reduces to a linear potential $|m_\chi-g\phi|n_\chi$ breaks down in a small region around $\phi=\phi_0$ where $\chi$ is relativistic. In this region, $V_{\rm eff}(\phi)$ is to leading order quadratic instead of linear in $\phi$. Nevertheless, this quadratic regime corresponds to a very small range of $\phi$ around $\phi=0$ and does not affect the dynamics of $\phi$ appreciably. We discuss this further in Appendix.~\ref{appendix:finitedensity}.} The quadratic minimum $\phi_*$ is relevant when $\phi<\phi_0$. It appears because adding the linear term  $-m_\phi^2\phi_*\phi$ to the quadratic potential $m_\phi^2\phi^2/2$ amounts to a shift in the quadratic potential.\footnote{ When $\phi<\phi_0$, the effective potential can be written as $V_{\rm eff}=m_\phi^2(\phi-\phi_*)^2/2-m_\phi^2\phi_*^2/2+m_\chi n_\chi$.} In the early universe, the ratio $\phi_*/\phi_0\propto n_\chi$ starts relatively high, likely greater than unity, in which case $\phi_0$ is the only minimum of $V_{\rm eff}(\phi)$. Once $\phi_*/\phi_0$ goes below unity, the minimum of $V_{\rm eff}$ switches to $\phi_*$, which moves progressively closer to the origin ($\phi=0$) due to its $\phi_*\propto n_\chi\propto a^{-3}$ scaling. Notice also that the values of $V_{\rm eff}$ at these minima are nonzero and given by $V_{\rm eff}(\phi_0)=m_\phi^2\phi_0^2/2$ and $V_{\rm eff}(\phi_*)=m_\chi n_\chi-m_\phi^2\phi_*^2/2$ where $-m_\phi^2\phi_*^2/2\propto a^{-6}$. As we will see, the constancy of $V_{\rm eff}(\phi_0)$ could lead to the dark sector as a whole, behaving as dark energy at the background level.

The quadratic piece of the potential dominates as long as $\phi \ll \phi_0$. Writing $\phi_0 = (4\pi G \alpha_{\chi\chi})^{-1/2}$ helps understand that a leading order perturbative expansion in $\alpha_{\chi\chi}$, such as the one performed in the previous precision cosmology studies of refs.~\cite{Bottaro:2024pcb,Bottaro:2023wkd,Archidiacono:2022iuu,Costa:2025kwt}, is tantamount to effectively sending $\phi_0 \rightarrow \infty$, or equivalently, saying that the mass shift $g\phi$ is never comparable to the bare mass scale $m_\chi$. While this is appropriate for the precision study of suitably small couplings, it cannot be assumed of regions of parameter space with large couplings.

\subsection{Scalar Evolution}

In our setting, the static Yukawa parametrization in Eq.~\eqref{eq:YukForce} corresponds to locking the cosmological background mediator field value $ \phi$ at the quadratic minimum $\phi_*$. However, as we will see, $ \phi$ in general deviates far from $\phi_*$ and instead evolves non-trivially according to its equation of motion
\begin{align}
\ddot{ \phi}+3H\dot{ \phi}+m_\phi^2[ \phi-\phi_*\text{sign}\left(\phi_0- \phi\right)]=0. \label{eq:EoM}
\end{align}
We solve this equation numerically for several representative parameters $\lambda_\phi$ and $\alpha_{\chi\chi}$, and plot the resulting magnitude of $ \phi$ as a function of scale factor $a$ in Fig.~\ref{fig:phievolution}. In solving Eq.~\eqref{eq:EoM}, we assume that the Hubble rate $H$ is given by the $\Lambda$CDM value. This is justified deep in RD, where the dark sector's energy density is subdominant, even if the background evolution of the dark sector deviates considerably from that in $\Lambda$CDM then.

Rather insensitively to the initial condition, the $\phi$ field initially tracks an attractor solution $\phi_{\rm rise}$, whose form during RD is
\begin{align}
    \phi_{\rm rise}=\frac{3\alpha_{\chi\chi}f_\chi}{4}\phi_0\frac{a}{a_{\rm eq}},\label{eq:phirise}
\end{align}
where $a_{\rm eq}^{-1}\approx 3400$ is the scale factor at matter-radiation equality. Physically, during this phase, where $\phi\ll \phi_0,\phi_*$, the $\phi$ field simply rolls down the initially dominant, linear potential piece $V_{\rm eff}\supset m_\phi^2\phi_*|\phi_0-\phi|$. After a sufficient field excursion, this rolling becomes insensitive to the initial values of $\phi$ and $\dot{\phi}$, asymptoting to the dynamical attractor $\phi_{\rm rise}$; we clarify the basin of attraction toward $\phi_{\rm rise}$ in Appendix.~\ref{appendix:scalardetail}. The existence of this attractor greatly reduces the sensitivity of our subsequent analyses to the initial conditions of $\phi$. This universal attractor solution $\phi_{\rm rise}$ is seen in the rising behavior on the left sides of all three of the red, green, and blue curves in Fig.~\ref{fig:phievolution}.

We show in Appendix~\ref{appendix:BBN} that as long as $\phi$ tracks the attractor solution  $\phi_{\rm rise}$ at early times, BBN does not place any constraint on $(\lambda_\phi,\alpha_{\chi\chi})$.

\begin{figure}[t!]
    \centering
    \includegraphics[width=\linewidth]{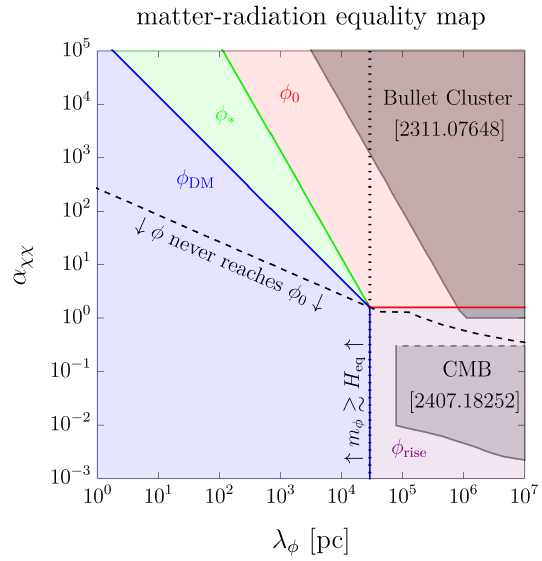}

    \caption{Types of solutions of $\phi$ at matter-radiation equality.  For example in the red region $\phi$ is at $\phi_0\equiv m_\chi/g$,  the linear finite-density minimum.  In the green region $\phi$ is at $\phi_*\equiv gn_\chi/m_\phi^2$, the quadratic finite-density minimum.  In the purple region it is at $\phi_{\rm rise}\equiv (3\alpha_{\chi\chi}f_\chi/4)\phi_0(a/a_{\rm eq)}$,  the attractor solution initially tracked by $\phi$ for a wide range of initial conditions, and in the blue region it is at $\phi_{\rm DM}$ which is a CDM-like solution where $\phi$ oscillates with its bare mass with an amplitude that scales as $a^{-3/2}$. In the red region, the dark sector behaves as dark energy to the left of the dotted line and as dark radiation to the right of the dotted line. Also shown are the regions ruled out by Bullet Cluster observation \cite{Graham:2024hah} and small-coupling analysis of the CMB data \cite{Bottaro:2024pcb}. The CMB limit has a ceiling because Ref.~\cite{Bottaro:2024pcb} employs a small-$\alpha_{\chi\chi}$ approximation in their analysis. We show a rough approximation to this ceiling as a dashed-gray upper boundary, corresponding to $\alpha_{\chi\chi}\sim 0.3$. Above (below) the black dashed line, $\phi$ does (does not) reach $\phi_0$, which corresponds to the large (small) mass-change regime, i.e. the $\phi_{\rm rise}\rightarrow\phi_0\rightarrow\phi_*\rightarrow\phi_{\rm DM}$ ($\phi_{\rm rise}\rightarrow\phi_{\rm DM}$) case.}
    \label{fig:eqmap}
\end{figure}

The tracking of the attractor $\phi_{\rm rise}$ continues until $\phi_{\rm rise}$ hits either the linear minimum $\phi_0$ or the quadratic minimum $\phi_*$. Depending on which of these comes first, $\phi$ subsequently oscillates around the linear minimum with a radiation-like equation of state ($\phi_0$ phase) or around the quadratic minimum with a CDM like equation of state ($\phi_{\rm DM}$ phase). This binary possibility causes the evolution of $\phi$ to branch into two cases:
\begin{enumerate}
    \item \textit{Small $\chi$ mass-change regime}:\\ 
    If $(3\alpha_{\chi\chi}f_\chi/4 )(\lambda_\phi H_{\rm eq})^{1/2}\lesssim 1$, $\phi$ goes through the following phases, $\phi_{\rm rise}\rightarrow\phi_{\rm DM}$, and so $\chi$'s effective mass remains $M_\chi\approx m_\chi$ throughout. This is shown in the blue curve in Fig.~\ref{fig:phievolution}. The increasing $\phi_{\rm rise}\propto a$ solution gets cut by the crossing with the decreasing $\phi_*\propto a^{-3}$. At that point, $\phi$ switches to oscillating around $\phi_*$ with its amplitude scaling as $\sim a^{-3/2}$ like that of an ultralight DM condensate \cite{Hu:2000ke,Ferreira:2020fam,Hui:2021tkt} until the present epoch.
    \item \textit{Large $\chi$ mass-change regime}:\\
    If $(3\alpha_{\chi\chi}f_\chi/4 )(\lambda_\phi H_{\rm eq})^{1/2}\gtrsim 1$, then $\phi$ goes through the following phases $\phi_{\rm rise}\rightarrow\phi_0\rightarrow\phi_*\rightarrow\phi_{\rm DM}$, in which case $M_\chi$ sometimes deviates significantly from $m_\chi$. This is shown in the red and green curves in Fig.~\ref{fig:phievolution}. The increasing $\phi_{\rm rise}\propto a$ first crosses $\phi_0$. Following that, $\phi$ oscillates around $\phi_0$ with its maximum speed scaling as $\dot{\phi}\sim H\phi\propto a^{-2}$, thus behaving at background level as dark radiation, $(\dot{\phi})^2/2\sim H^2{\phi}^2\propto a^{-4}$. Once the Hubble $H$ goes below the bare mass $m_\phi$, the energy density of the constant mass term $m_\phi^2\phi_0^2/2$ becomes dominant over the kinetic energy of $\phi$, and the dark sector subsequently behaves as dark energy. This continues until $\phi_0$  is crossed by $\phi_*\propto a^{-3}$. Then, $\phi$ turns to oscillating around $\phi_*$. Because the oscillation of $\phi$ around $\phi_0$ had been significantly Hubble damped between the $\phi_{\rm rise}\rightarrow \phi_0$ and $\phi_0\rightarrow \phi_*$ points, the oscillation around $\phi_*$ is at first tiny compared to $\phi_*$, and so $\phi$ appears to closely track $\phi_*$. After a while, as the oscillation amplitude scales as $\propto a^{-3/2}$, which is slower than $\phi_*\propto a^{-3}$, this oscillation eventually dominates, and from then on $\phi$ behaves as an ultralight DM condensate with an amplitude scaling as $\propto a^{-3/2}$. 
\end{enumerate}
The case boundary $(3\alpha_{\chi\chi}f_\chi/4 )(\lambda_\phi H_{\rm eq})^{1/2}\sim 1$ applies only for $\lambda_\phi\lesssim H_{\rm eq}^{-1}$, in which case the $\phi_{\rm rise}$ does not extend beyond RD. For $\lambda_\phi\gg H_{\rm eq}^{-1}$, this boundary asymptotes to $\alpha_{\chi\chi}f_\chi\sim 0.1$.

Depending on the place in parameter space $(\lambda_\phi,\alpha_{\chi\chi})$,  at matter-radiation equality the $\phi$ field can be tracking $\phi_{\rm rise}$, tracking $\phi_0$, oscillating around $\phi_0$ (with dark-radiation like scaling), tracking $\phi_*$, or oscillating around $\phi_*$ (with CDM-like scaling). These possibilities are mapped in Fig.~\ref{fig:eqmap}. Given that the cosmic history is particularly well measured at $z\sim 10^3$, around matter-radiation equality, we can already deduce that parameter space points that yield $\phi\approx \phi_0$ then are almost certainly ruled out, as they predict a dark sector that behaves very differently from CDM. Less obviously, we will see in the next subsection that parameter space that yields $\phi\approx\phi_0$ prior to that, at $z\sim 10^{3}-10^{5}$, may also be ruled out when confronted with cosmological data which have some sensitivity to this earlier epoch. Moreover, we have some probes of cosmic-perturbation modes that entered the horizon at even earlier times. As we will show in Section.~\ref{s:perturbations}, these can rule out additional parameter space that predicts a dark sector deviating significantly from CDM at  $z\gtrsim 10^{5}$.

We also show in Fig.~\ref{fig:eqmap} the existing limits from Bullet Cluster observation \cite{Bogorad:2023wzn} and Ref.~\cite{Bottaro:2024pcb}'s analysis of Planck CMB data \cite{Planck:2018vyg}. Incidentally, the Bullet Cluster limit lies almost entirely in the the $\phi_0$ regime, whereas the small-$\alpha_{\chi\chi}$ CMB limit lies completely in the $\phi_{\rm rise}$ regime. In obtaining their limits, Ref.~\cite{Bottaro:2024pcb} employs a small-$\alpha_{\chi\chi}$ approximation in their analysis. Thus this analysis breaks down at large enough $\alpha_{\chi\chi}$.  
For instance, a nonzero $\alpha_{\chi\chi}$ leads to a shift $g\phi$ in the effective mass of $\chi$. The small-coupling treatment requires this mass shift stays small compared with the bare mass, $g\phi\ll m_\chi$, so that $\phi$ never reaches the linear minimum $\phi_0=m_\chi/g$, where the dark sector deviates starkly from CDM. To roughly represent the $\alpha_{\chi\chi}$ cutoff above which their analysis is expected to be invalid, we draw a gray-dashed line at the top of the ruled-out region at $\alpha_{\chi\chi}\sim 0.3$. Note that this cutoff is not exact; its precise value is not to be taken literally. 

Naively, as we increase the self-interaction of DM from zero, we would expect its behavior to deviate increasingly from that of CDM. However, it is possible that beyond the small-coupling regime, entirely new dynamics takes over and reverses this tendency, making the cosmology more viable. For instance, one could imagine in some models that when the DM self-interaction is increased beyond a certain threshold the dark sector could clump into compact macroscopic objects whose large-scale behavior returns to resembling CDM. Therefore, small-coupling analyses that rule out couplings larger than a certain value should end their exclusion region at the largest couplings where their small-coupling approximation is valid. Moreover, existing data may be re-interpreted through the lens of different or less-minimal dark sector models, where the boundaries that delineate what is ruled out and what is not could change. It is therefore important to understand the actual physical effects that rule out a given parameter space.

\subsection{Catastrophic Dark Sector Equation of State}

The background energy density $\rho_{\rm dark}$ and pressure $p_{\rm dark}$ of the  $\phi$ and $\chi$ fields, collectively, are given by
\begin{align}
    \rho_{\rm dark}&=|m_\chi-g\phi|n_\chi+\frac{1}{2}\dot{\phi}^2+\frac{1}{2}m_\phi^2\phi^2\label{eq:rhodark},\\
    p_{\rm dark}&=\frac{1}{2}\dot{\phi}^2-\frac{1}{2}m_\phi^2\phi^2\label{eq:pdark},
\end{align}
where we have neglected the contribution to $p_{\rm dark}$ from $\chi$ which we assume to be cold. On timescales longer than that of $\phi$ oscillation, the dark sector energy density scales as $\rho_{\rm dark}\propto a^{-3(\bar{w}_{\rm dark}+1)}$, with an average equation of state $\bar{w}_{\rm dark}$ given by
\begin{align}
    \bar{w}_{\rm dark}&=\frac{\left<\dot{\phi}^2/2-m_\phi^2\phi^2/2\right>}{\left<|m_\chi-g\phi|n_\chi+\dot{\phi}^2/2+m_\phi^2\phi^2/2\right>}\nonumber\\
    &=\begin{cases}
        0, &\phi_{\rm rise} \text{ or osc. around }\phi_* \,\,(|\phi|\ll \phi_0)\\
        1/3, &\text{osc. around }\phi_0\,\,(|\phi-\phi_0|\gg \phi_0)\\
        -1, &\text{tracking }\phi\approx \phi_0\,\,(|\phi-\phi_0|\ll \phi_0),
    \end{cases}\label{eq:wdark}
\end{align}
which corresponds to $\rho_{\rm dark}\propto a^{-3},a^{-4},a^0$, respectively. Here, angled brackets $\left<\ldots\right>=\int_0^{T_{\rm osc}}dt\ldots/T_{\rm osc}$ denote a time average over an oscillation period $T_{\rm osc}$, assumed $\ll H^{-1}$. For an oscillation around a polynomial potential $V(\phi)\propto \phi^n$, the time average of the kinetic energy of $\phi$ evaluates to $\left<\dot{\phi}^2/2\right>=n\left<V\right>/2$ as per virial theorem.  The evolution regimes where the dynamics of $\phi$ is mainly dictated by the linear potential $|m_\chi-g\phi|n_\chi$ and by the quadratic potential $m_\phi^2\phi^2/2$ correspond to $n=1$ and $n=2$, respectively.

According to Eq.~\eqref{eq:wdark}, the $\bar{w}_{\rm dark}$ deviates significantly from zero when $\phi$ is either oscillating around or closely tracking $\phi_0$. Specifically, our estimates (detailed in Appendix~\ref{appendix:scalardetail}) show that it is radiation-like ($\bar{w}_{\rm dark}\approx 1/3$) from $a=(3\alpha_{\chi\chi} f_\chi/4)^{-1}a_{\rm eq}$ to $H\sim m_\phi$ and dark-energy-like ($w_{\rm dark}\approx -1$) from $H\sim m_\phi$ to $a=(3\alpha_{\chi\chi} f_\chi/4)^{1/3}(\lambda_\phi H_{\rm eq})^{2/3}a_{\rm eq}$. Such a large deviation of $\bar{w}_{\rm dark}$ from the CDM-like $\bar{w}_{\rm dark}\approx 0$ only occurs in the large mass-change regime ($\phi_{\rm rise}\rightarrow \phi_0\rightarrow \phi_{*}\rightarrow\phi_{\rm DM}$).  In the small mass-change regime ($\phi_{\rm rise}\rightarrow \phi_{\rm DM}$), in which $\phi$ does not reach $\phi_0$, the average $w_{\rm dark}$ remains CDM-like throughout. Thus, we expect cosmological observations to yield strong limits particularly in the large mass-change case.

\begin{figure}[t!]
    \centering
    \includegraphics[width=\linewidth]{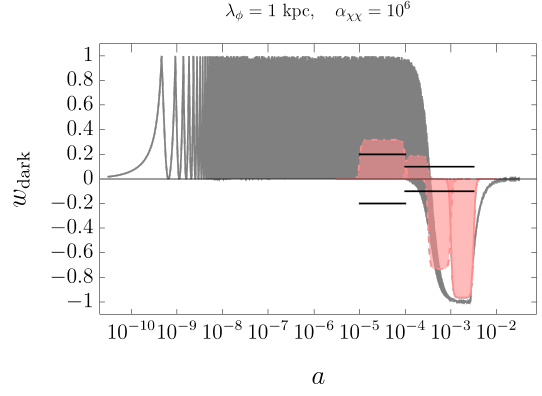}\\
    \caption{Evolution of the instantaneous dark sector equation of state $w_{\rm dark}=p_{\rm dark}/\rho_{\rm dark}$ (solid gray) as a function of the scale factor $a$. The dark sector pressure $p_{\rm dark}$ and energy density $\rho_{\rm dark}$ are defined in Eqs.~\eqref{eq:rhodark}\&\eqref{eq:pdark}. The widths and amplitudes of the red bars represent the four scale-factor bins 1, 2, 3, 4 (see Eq.~\eqref{eq:abins}) used in our analysis and the averaged equation of state $\bar{w}_{\rm dark}^{(i)}$ within them. The black lines mark the boundaries of the allowed ranges of $\bar{w}_{\rm dark}^{(i)}$; we consider parameter space with red bars protruding the black lines as ruled out.}
    \label{fig:wdark}
\end{figure}

We plot in Fig.~\ref{fig:wdark} the evolution of the instantaneous $w_{\rm dark}$ as a function of the scale factor $a$ for the parameters $\lambda_\phi=10\text{ kpc}$ and $\alpha_{\chi\chi}=100$, representative of the large mass-change case ($\phi_{\rm rise}\rightarrow \phi_0\rightarrow \phi_{*}\rightarrow \phi_{\rm DM}$). During most of $\phi_{\rm rise}$, the energy density of $\phi$ is negligible and $w_{\rm dark}\approx 0$. $w_{\rm dark}$ begins to rise when $\phi_{\rm rise}$ approaches $\phi_0$. Once $\phi$ switches to oscillating under the linear potential around $\phi_0$, $w_{\rm dark}$ becomes spiky, varying between $0$ and $1$, and averages to $1/3$. This is seen in Fig.~\ref{fig:wdark} in the rapid oscillations of the gray line between $a\sim 10^{-9}$ and $a\sim10^{-4}$. After the $\phi$ oscillation gets  sufficiently Hubble damped, the value of $\phi$ essentially freezes at $\phi_0$, and the bare mass term $m_\phi^2\phi_0^2/2$ behaves as dark energy, with $\bar{w}_{\rm dark}\approx -1$. This is seen when the gray line is near -1 around $a \sim 10^{-3}$. At some point, $\phi_*$ crosses $\phi_0$, whereupon $\phi$ begins oscillating around the moving $\phi_*$. As that happens, $w_{\rm dark}$ departs from -1 and begins rising, while $\phi$ closely tracks $\phi_*$. Once the amplitude of $\phi$'s oscillation around $\phi_*$ becomes large compared to $\phi_*$, the $\phi$ field starts behaving as CDM, and $w_{\rm dark}$ then undergoes tiny oscillations around zero such that on average $\bar{w}_{\rm dark}\approx 0$.

\subsection{Constraints on Dark Matter Equation of State}
\label{section:EoSconstraints}

Refs.~\cite{Ilic:2020onu, Kopp:2018zxp} have performed relatively model-independent analyses to place constraints on the evolution of the dark sector's equation of state $w_{\rm dark}(a)$ across the observable cosmic history. Instead of assuming a specific functional form of $w_{\rm dark}(a)$, as commonly done, they allow its values at different epochs to float freely in bins of scale factor.\footnote{ See also Refs.~\cite{2012PhRvD..86l3504S,Hojjati:2013oya}, which are similar in spirit. These works constrain the Hubble rate at different epochs, instead of the dark sector equation of state. However, they only allow \textit{positive} deviations in Hubble relative to the $\Lambda$CDM evolution. On the other hand, our scenario actually predicts local-in-time Hubble deficits relative to $\Lambda$CDM, and so their analyses unfortunately do not cover our scenario.} The equations of state in the different bins then serve as extra parameters, in addition to the six standard $\Lambda$CDM parameters, with which to fit cosmological data, such as those of CMB anisotropies. Because the $w_{\rm dark}(a)$ assumed in their analyses are allowed to vary in a less-constrained and model-independent manner, their analyses lead to constraints that are not only conservative but also applicable to a wide range of scenarios, including ours.

Ref.~\cite{Ilic:2020onu} divided the scale factor range from $10^{-5}$ to $1$ into eight logarithmic scale-factor bins. We consider only the first four  of these bins, corresponding to the scale factor ranges:  
\begin{align}
    \underbrace{[10^{-5}, 10^{-4}]}_{\text{bin 1}}, \underbrace{[10^{-4},10^{-3.5}]}_{\text{bin 2}}, \underbrace{[10^{-3.5},10^{-3}]}_{\text{bin 3}}, \underbrace{[10^{-3},10^{-2.5}]}_{\text{bin 4}}.\label{eq:abins}
\end{align}
These are the most constraining bins for the model because it behaves very closely to $\Lambda$CDM at scale factors later than these bins. We denote the average equation of state in bin $i=1,2,3,4$ as 
\begin{align}
    \bar{w}_{\rm dark}^{(i)}\equiv \frac{\int_{\text{bin }i} d\ln a \,w_{\rm dark}}{\int_{\text{bin }i} d\ln a}.
\end{align}
The results of ref.~\cite{Ilic:2020onu} show that consistency with CMB power spectrum (Planck 2015), CMB lensing likelihood (Planck 2015), BAO (6dF galaxy survey and SDDS-III BOSS), and the Hubble Space Telescope prior on $H_0$, restricts the $\bar{w}_i$s to certain ranges close to zero. In some bins close to matter-radiation equality, their limits are so stringent that they even ruled out $\Lambda$CDM at 99\% confidence level. While our treatment is inspired by the analysis of ref.~\cite{Ilic:2020onu}, to keep our results robust, we instead consider the following very conservative approximations to their limits 
\begin{align}
    -0.2<\bar{w}_{\rm dark}^{(1)}<+0.2,\quad -0.1<\bar{w}_{\rm dark}^{(2,3,4)}<+0.1 .\label{eq:wlimits}
\end{align}
These are shown as black solid lines in Fig.~\ref{fig:wdark}. We refer the reader to Fig.~\ref{fig:wlimitcomparison} in Appendix.~\ref{appendix:extrafigures} for a comparison between the actual ranges of allowed $\bar{w}_i$ found in ref.~\cite{Ilic:2020onu} and the very conservative ones we assume. Since the $w_{\rm dark}$ in our scenario can, at times, deviates far from zero (see Fig.~\ref{fig:wdark}), the resulting limits on dark forces do not depend sensitively on the precise values of the maximum and minimum allowed $\bar{w}_{\rm dark}^{(i)}$ assumed, as long as they are well below  $1/3$ and well above $-1$, respectively. We discuss the dark-force constraints that result from imposing Eq.~\eqref{eq:wlimits} in Section~\ref{s:results}. The results are shown in Fig.~\ref{fig:results}.

We note that any constraints on background cosmology based on probes of cosmic perturbations, e.g. the CMB power spectra, are unavoidably intertwined with details of the perturbation evolution, which we do not study in full in this work (we only consider a limited regime in Section.~\ref{s:perturbations} where the perturbations behave relatively simply). Ref.~\cite{Ilic:2020onu} takes a step further to make their analysis less sensitive to assumptions on cosmic perturbation evolution. They adopt the generalized DM parametrization \cite{Hu:1998kj}, which in addition to $w_{\rm dark}$ includes also the dark sector's sound speed $c_{s}$ and shear viscosity $c_{\text{vis}}$ for each scale-factor bin, both allowed to float, though assumed wavenumber-independent. After marginalizing over the $c_{s}$ and $c_{\text{vis}}$, ref.~\cite{Ilic:2020onu} found constraints on $\bar{w}_{\rm dark}$ that are less stringent but comparable to that of ref.~\cite{Kopp:2018zxp}, which assumes $c_s=c_{\rm vis}=0$. This greatly reduces but does not completely remove the sensitivity of the results to unmodeled cosmic perturbations. In fact, in some cases of our interest, the evolution of the dark sector's sound speed is not only highly non-trivial (and so may not be well-represented by coarse-grained values) but also wavenumber-dependent.  Nevertheless, achieving an exact cancellation between the effects of dark-sector perturbations and those of a modified background expansion would require an extraordinary degree of fine-tuning. We therefore regard such scenarios as highly implausible, and consequently exclude regions of parameter space that produce extreme departures from the standard background evolution violating Eq.~\eqref{eq:wlimits}.

\section{Cosmic Perturbations}
\label{s:perturbations}

Here, we study the evolution of over-densities in DM in the presence of dark forces. Dark forces can have drastic consequences on cosmic perturbations even in regimes where the background evolution is consistent with $\Lambda$CDM. Hence, cosmic-perturbation considerations can place limits complementary to those obtained from background cosmology.

\subsection{Meszaros-Like Perturbation Equation }

It has long been suggested that dark forces can induce exponential growth in DM perturbations \cite{Afshordi:2005ym,Domenech:2023afs,Savastano:2019zpr}. Given the catastrophic outcome it may cause, we will focus on this effect in constraining dark forces through perturbation analyses. We note that the exponential growth is based on the assumption that the effect of coupling the DM $\chi$ to the mediator $\phi$ amounts to replacing the Newton's constant $G$ with $G_{\rm eff}(k)=G\left[1+\alpha_{\chi\chi} k^2/(a^2m_\phi^2+k^2)\right]$ in cosmic perturbation equations, namely the Meszaros equation. This corresponds to considering only the particular solutions to the mediator's equation of motion Eq.~\eqref{eq:EoM} and its perturbation, which physically can be interpreted as the mediator field being sourced in a quasi-static manner by the DM particles, as in electrostatics. In Appendix~\ref{appendix:perturbation}, we derive the Meszaros-like perturbation equation with $G\rightarrow G_{\rm eff}$ (Eq.~\eqref{eq:Meszaroslike}) from first principles, clarifying under which conditions this equation is valid. We find that the cosmic perturbation equation only reduces to such a form when $\phi$ is closely tracking its quadratic finite-density minimum $\phi_*=gn_\chi/m_\phi^2$ while its value is sufficiently small $\phi_*\ll\phi_0$.

We have seen in Section~\ref{s:background} that the mediator field $\phi$ has rich background dynamics where it generically deviates far from $\phi_*$. While this suggests that the substitution $G\rightarrow G_{\rm eff}$ is rarely valid, we have sufficient understanding of $\phi$'s background evolution to pinpoint when $\phi\approx \phi_*$ is actually satisfied. As found in Section~\ref{s:background}, this occurs only in the large mass-change case ($\phi_{\rm rise}\rightarrow \phi_0\rightarrow \phi_*\rightarrow \phi_{\rm DM}$), starting right after $\phi_0$ is crossed by $\phi_*$, at scale factor $a_{\phi_0=\phi_*}$, and ending when the $\phi$ background begins to redshift as CDM at scale factor $a_{\phi_*=\phi_{\rm DM}}$. For example, in Fig.~\ref{fig:phievolution} the tracking of $\phi_*$ occurs for the green curve at $a/a_{\rm eq}\sim 1$ and for the red curve at $a/a_{\rm eq}\gtrsim 10$. In terms of the model parameters, $a_{\phi_0=\phi_*}$ and $a_{\phi_*=\phi_{\rm DM}}$ are given by (see Appendix.~\ref{appendix:scalardetail})
\begin{align}
    a_{\phi_0=\phi_*}&=\left(\frac{3\alpha_{\chi\chi} f_\chi}{4}\right)^{1/3}\left(\frac{\lambda_\phi}{H_{\rm eq}^{-1}}\right)^{2/3}a_{\rm eq},\\
    a_{\phi_*=\phi_{\rm DM}}&=\left(\frac{3\alpha_{\chi\chi} f_\chi}{4}\right)^{7/9}\left(\frac{\lambda_\phi}{H_{\rm eq}^{-1}}\right)^{8/9}a_{\rm eq}.
\end{align}

In order to approximately capture the exponential growth of the DM density contrast $\delta_\chi\equiv\delta n_\chi/n_\chi$, we consider the following Meszaros-like equation \cite{1974A&A....37..225M}
\begin{align}
    \ddot{\delta}_\chi+2H\dot{\delta}_\chi&=\frac{3H^2}{2}\Omega_\chi\delta_\chi\left[1+\frac{\alpha_{\chi\chi }\Theta_*(t)}{1+\left(\frac{a/k}{\lambda_\phi}\right)^{2}}\right],\label{eq:Meszaroslike}
\end{align}
where we have defined
\begin{align}
    \Theta_*(t)&=\Theta[a(t)-2a_{\phi_0=\phi_*}]\Theta[a_{\phi_*=\phi_{\rm DM}}-a(t)].\label{eq:phistarwindow}
\end{align}
and $\Omega_\chi=m_\chi n_\chi/\rho_{\rm crit}$, $\Theta$ without a subscript is the Heaviside step function, and we have defined $\Theta_*$ such that it is equal to unity when $\phi$ is closely tracking $\phi_*$ and zero otherwise. We assume that $H$ is given by the $\Lambda$CDM value. We start the support at $a=2a_{\phi_0=\phi_*}$ instead of at $a=a_{\phi_0=\phi_*}$ in order to ensure $\phi_*\ll \phi_0$ and that $w_{\rm dark}$ and its time derivative are negligible (which also implies $|\dot{M}_\chi/M_\chi|\ll H$). Enhanced growth of $\delta_\chi$ occurs when the right-hand side of Eq.~\eqref{eq:Meszaroslike} is $\gtrsim H^2\delta_\chi$. If at some point during the $\phi\approx \phi_*$ tracking phase the physical wavenumber enters the range of the dark force,  $a/k\ll \lambda_\phi$, then it suffices to have $\alpha_{\chi\chi}\gtrsim 1$. In the opposite case, when $a/k\gg \lambda_\phi$, the $\alpha_{\chi\chi}$ needs to be large enough that $\Omega_{\chi}\alpha_{\chi\chi}(\lambda_\phi k/a)^2\gtrsim 1$.

\begin{figure}[t!]
    \centering
    \includegraphics[width=\linewidth]{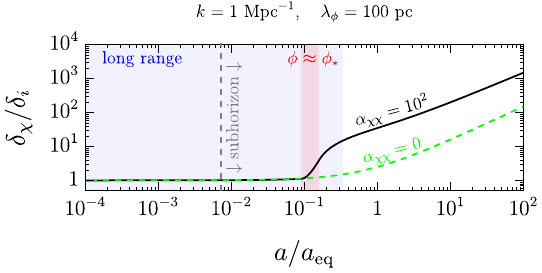}
    \includegraphics[width=\linewidth]{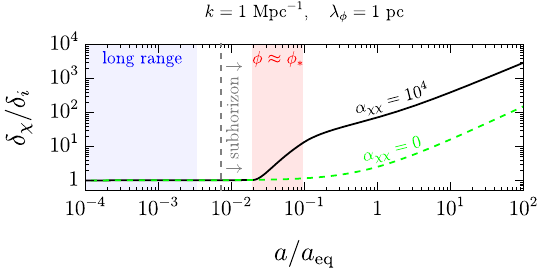}
    \caption{The evolution of the DM density contrast $\delta_\chi=\delta n_\chi/n_\chi$ (solid black line, normalized to its initial, superhorizon value) as a function of the scale factor $a$ (normalized to that at matter-radiation equality, $a_{\rm eq}$). Here, we follow the mode with comoving wavenumber $k=1\text{ Mpc}^{-1}$, and set the dark-force parameters as indicated in the plots. The gray dashed line marks the point where this mode becomes subhorizon. The black solid lines represent the evolution of $\delta_\chi$ as dictated by Eq.~\eqref{eq:Meszaroslike}, to be contrasted with the evolution in the absence of dark forces ($\alpha_{\chi\chi}=0$) as given by Eq.~\eqref{eq:deltaLambdaCDM} and shown in dashed green lines. The light red region shows the region where $\phi$ is expected to closely track $\phi_*$, in which the $\Theta_*$ in Eqs.~\eqref{eq:Meszaroslike} and \eqref{eq:phistarwindow} is equal to unity.  The light blue region indicates where the dark-force range is longer than the physical wavelength $a/k$ of the mode. The \textit{top} (\textit{bottom}) plot corresponds to the case where the dark force is long-range $\lambda_\phi\gg a/k$ (short-range $\lambda_\phi\ll a/k$) when $\phi\approx \phi_*$. It can be seen that enhanced growth occurs in the light red region in either case. Furthermore, $\delta_\chi$ appears to continue growing relative to the $\Lambda$CDM one even outside of the light red region. This is due to the speed $\dot{\delta}_\chi$ acquired in the light red region.}
    \label{fig:perturbationgrowth}
\end{figure}

We plot in Fig.~\ref{fig:perturbationgrowth} the evolution of $\delta_\chi$ in the presence of dark forces relative to that in $\Lambda$CDM. A dramatic, exponential increase in the growth of $\delta_\chi$ is clearly visible. The figure also shows that to cause an exponential growth, the dark force need not be long-range compared to the physical wavelength of the mode. Interestingly, $\delta_\chi$ continues to grow even outside the region where $\Theta_*=1$. This is due to a combination of the inertia of $\delta_\chi$, as indicated by the presence of $\ddot{\delta}_\chi$ term in Eq.~\eqref{eq:Meszaroslike}, and the speed $\dot{\delta}_\chi$ acquired during the time window when $\Theta_*=1$.

\begin{figure}
    \centering
    \includegraphics[width=\linewidth]{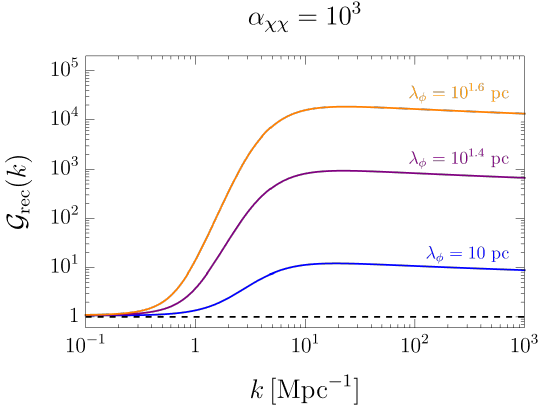}        \caption{Extra growth of DM density contrast $\mathcal{G}_{\rm rec}(k)=[\delta_\chi(k)/\delta_{\chi}^{\Lambda\text{CDM}}]_{\rm rec}$ by the epoch of recombination as a function of comoving wavenumber $k$. Here, the $\delta_\chi$ and $\delta_\chi^{\Lambda\text{CDM}}$ are evolved according to Eq.~\eqref{eq:Meszaroslike} and Eq.~\eqref{eq:deltaLambdaCDM}, respectively, with the dark-force parameters as indicated in the plot.}
    \label{fig:growthspectrum}
\end{figure}

Note that the density contrast $\delta_\chi$ grows considerably even in $\Lambda$CDM, in the absence of dark forces. In order to quantify the enhancement due to dark forces relative to the $\Lambda$CDM case (corresponding to $\alpha_{\chi\chi}=0$), we define the following quantity
\begin{align}
    \mathcal{G}_{\rm rec}(k)\equiv \left.\frac{\delta_\chi(k)}{\delta_\chi^{\Lambda\text{CDM}}(k)}\right|_{ a=a_{\rm rec}},\label{eq:relgrowthfactor}
\end{align}
which we will refer to as the \textit{relative growth factor}. We have chosen to evaluate this ratio at the epoch of recombination, $a_{\rm rec}\approx (1100)^{-1}$, for practical reasons which will become clear later. Comoving modes with $k\gg k_{\rm eq}\approx 0.01\text{ Mpc}^{-1}$ cross the horizon far before matter-radiation equality, at $a_k\approx (k_{\rm eq}/\sqrt{2}k)a_{\rm eq}$. For such modes and $a\gtrsim a_{\rm eq}$, we can approximate the DM density contrast in $\Lambda$CDM as  \cite{Hu:1995en}
\begin{align}
    \delta_\chi^{\Lambda\text{CDM}}(k)\approx  \frac{3}{5}\frac{k^2}{\Omega_mH_0^2}\Phi_p(k)T(k)D(a)\label{eq:deltaLambdaCDM},
\end{align}
where the transfer function $T(k)$ and growth function $D(a)$ can be found in \cite{Hu:1995en}. We assume $\Phi_p$ to be the usual adiabatic, scale-invariant, superhorizon perturbation, and take as fiducial values $\Omega_m=0.27$ and $H_0=70\text{ km}/\text{s}/\text{Mpc}$. Fig.~\ref{fig:growthspectrum} shows how the extra growth due to long-range forces distorts the DM density contrast spectrum as quantified by $\mathcal{G}_{\rm rec}(k)$. In  $k$-space, the dark-force induced enhancement starts at $k\sim \alpha_{\chi\chi}^{-1/3}(\lambda_\phi H_{\rm eq})^{-2/3}k_{\rm eq}$ which corresponds to $\Omega_{\chi}\alpha_{\chi\chi}(\lambda_\phi k/a)^2\sim 1$ at $a_{\phi_0= \phi_*}$. The enhancement plateaus (becomes nearly $k$-independent) when $k$ exceeds $k\sim\alpha_{\chi\chi}^{7/9}(\lambda_\phi H_{\rm eq})^{-1/9}$, in which case the physical size of the mode stays smaller than $\lambda_\phi$ throughout the $\phi\approx\phi_*$ phase, and consequently the evolution of $\delta_\chi$ becomes $k$-independent. This happens when $a_{k\lambda}=k\lambda_\phi$ becomes larger than $a_{\phi_*=\phi_{\rm DM}}$.

A possible concern is that the aforementioned, approximate scale-invariant enhancement in $\delta_\chi$ at high wavenumbers is such a simple effect that it can easily be degenerate with other effects, such as a compensating change to the primordial perturbation amplitude. Nevertheless, requiring that $\phi$ begins tracking $\phi_*$ before recombination, $a_{\phi_0=\phi_*}\lesssim a_{\rm rec}$ implies that the wavenumber above which there is enhancement, $k\sim \alpha_{\chi\chi}^{-1/3}(\lambda_\phi H_{\rm eq})^{-2/3}k_{\rm eq}$, is higher than $ k_{\rm eq}$. In other words, the turning-on of the enhancement occurs at some wavenumber $k\gtrsim k_{\rm eq}$, which generically lies within the range of wavenumbers for which we have probes. Hence, the enhanced clustering due to the dark force can manifest as a non-trivial shape-altering effect on the observable power spectrum, and not just an increase in the overall amplitude of the power spectrum.

\subsection{Scale-Dependent Growth Enhancement of Cosmic-Perturbations}

In $\Lambda$CDM cosmology, modes that become subhorizon during RD typically grow by at most a factor of $\mathcal{O}(10)$ before the epoch of CMB decoupling. By contrast, we have seen that dark forces can cause orders of magnitude of extra growth in $\delta_\chi$, during (and also slightly after) the scale-factor window where background field $\phi$ is expected to closely track $\phi_*$. We expect such a drastic modification to the evolution of $\delta_\chi$ to have clearly observable implications in cosmological data.

The clustering of DM is imprinted in the linear matter power spectrum, which is particularly well probed for wavenumbers $k\sim 10^{-3}-10\text{ Mpc}^{-1}$ by CMB anisotropy, galaxy surveys, weak lensing, and Lyman-$\alpha$ forest. At higher $k$'s matter perturbations have been relatively poorly probed. Recently, constraints on higher-$k$ modes are strengthening \cite{Graham:2024hah,Graham:2023unf,Graham:2025fdt,Iovino:2024tyg,Bringmann:2025cht}. In particular, observations of ultrafaint dwarf galaxies (UFDs) could be used to place constraints on the abundance of DM substructures expected from enhanced small-scale DM perturbations. Despite being relatively less-stringent in terms of sensitivity to the amplitude of $\delta_\chi$ (or equivalently the matter power spectrum), these higher-$k$ constraints from UFDs are actually far better at probing the dark-force parameter space than the more-stringent, lower-$k$ probes, since dark-force-induced growths tend to add orders of magnitude of power at higher wavenumbers $k\gg k_{\rm eq}$.

Based on the constraints from Lyman-$\alpha$ and UFDs, as reported in \cite{Graham:2024hah}, we consider the parameter space satisfying the following conditions (significantly weakened to be conservative) as ruled out:
\begin{align}
    \mathcal{G}_{\rm rec}(k=2\text{ Mpc}^{-1})&\gtrsim 10 &&\text{(Lyman-$\alpha$)},\nonumber\\
    \mathcal{G}_{\rm rec}(k=300\text{ Mpc}^{-1})&\gtrsim 10^2 &&\text{(UFD heating)},\label{eq:GrowthThresholds}
 \end{align}
where  $k= 2\text{ Mpc}^{-1}$ and $k=300\text{ Mpc}^{-1}$ represent, respectively, the highest wavenumber $k$ probed by Lyman-$\alpha$ and the $k$ for which UFD is most sensitive. We will discuss the constraints that result from imposing these conditions in Section~\ref{s:results}. Note that we have deliberately chosen highly conservative  $G_{\rm rec}(k)$ thresholds in Eq.~\eqref{eq:GrowthThresholds} in order to make our results more robust. We have checked that the final constraints are insensitive to the exact threshold values used due to the rapid rate at which $G_{\rm rec}(k)$ increases as $\alpha_{\chi\chi}$ and $\lambda_\phi$ are varied near the approximate boundaries of the constrained regime.

Apart from Lyman-$\alpha$ and UFDs, FIRAS's null measurements of CMB spectral distortions should also place some constraints on extra DM perturbation growths. Spectral distortions can arise from Silk damping of small-scale baryon-photon fluid perturbations which leads to inhomogeneous heating of the primordial plasma. Enhanced DM perturbations affect the pre-damping baryon-photon fluid perturbations through their contributions to the Newtonian potential $\Phi$. The damping of a given baryon-photon fluid mode occurs primarily at the moment it becomes smaller than the (time-dependent) damping scale $k_d$, typically far before matter-radiation equality. Hence, we expect the impact of enhanced DM perturbations on $\Phi$ to suffer some suppression due to both the subdominance of the DM energy density and the high pressure of the baryon-photon fluid. Nevertheless, capturing these effects properly would require solving the full, coupled Boltzmann and Einstein equations, which we leave for future work.

\begin{figure*}[t!]
    \centering
   \includegraphics[width=0.49\linewidth]{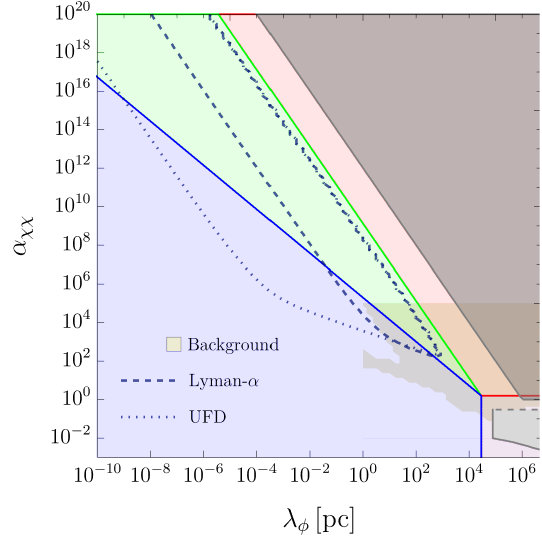}
    \includegraphics[width=0.49\linewidth]{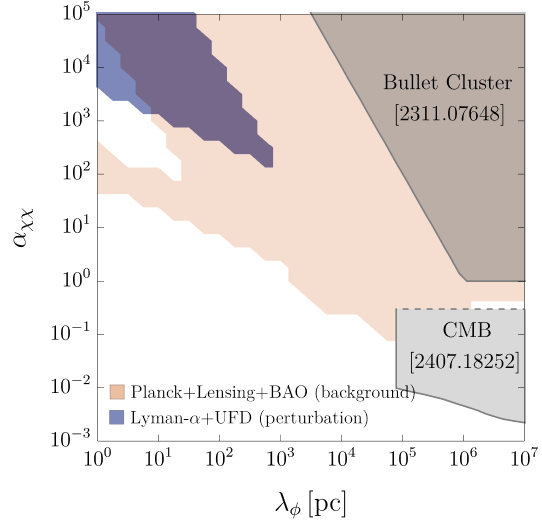}
    \caption{\textit{Right:} Combined limits on dark forces based on cosmological background analysis (beige) and perturbation analysis (dark blue). We also show existing limits from Bullet Cluster observation \cite{Bogorad:2023wzn} and Ref.~\cite{Bottaro:2024pcb}'s analysis of CMB data. The CMB limit has a ceiling because Ref.~\cite{Bottaro:2024pcb} employs a small-$\alpha_{\chi\chi}$ approximation in their analysis. We indicate the rough upper boundary of the regime of validity of their result with dashed-gray. We do not claim that the beige region continues to the left (small $\lambda_\phi$ direction) and upward (large $\alpha_{\chi\chi}$  direction) of the plot, although in principle it could. We stop our background analysis at $\lambda_\phi=1\text{ pc}$ because it becomes too computationally costly at $\lambda_\phi\ll 1\text{ pc}$. However, we expect the left edge of the beige region (apart from the part that sticks out, which we discuss in Section.~\ref{ss:mixedDM}) to be covered by the dark blue region, which does extend to smaller $\lambda_\phi$. \textit{Left: }The same dark-force limit from perturbation analysis as in the right plot (where it was shown in dark blue), but extended to much smaller $\lambda_\phi$ and much larger $\alpha_{\chi\chi}$ to illustrate the extent of the limit. The individual limits from the two probes that make up the totality of perturbation limit in the right figure are shown in the left figure in dashed (Lyman-$\alpha$) and dotted (UFD) lines. The light blue, light green, and light red colored regions correspond to that of the matter-radiation equality map of Fig.~\ref{fig:eqmap}. For comparison, the same beige region in the right plot is reproduced in the left plot, where now its perceived color changes due to being overlaid by the light blue, green, and red regions.}
    \label{fig:results}
\end{figure*}

\section{Results}
\label{s:results}

\subsection{Limits from Background Cosmology}
\label{ss:backgroundresults}
\subsubsection{Fixed $\chi$ Abundance}
The dark-force constraint that results from our cosmological-background analysis is shown on the right panel of Fig.~\ref{fig:results} in beige color.  Here, we set $f_\chi=f_{\rm DM}=0.85$ and solve Eq.~\eqref{eq:EoM} numerically with the initial conditions $\phi=0$ and $\dot{\phi}=0$ at the scale factor $a=10^{-9}$. We then use the solution to compute the average equation of state $\bar{w}_{\rm dark}^{(i)}$ in each bin, which we then confront  with the allowed ranges of $\bar{w}_{\rm dark}$ we assumed in Eq.~\eqref{eq:wlimits}. Fig~\ref{fig:results} shows that the constrained regime lies primarily in the large mass-change part of the parameter space (see Fig.~\ref{fig:eqmap}), as expected. For the most part, the constraint arises from the fact that when $\phi$ is oscillating around $\phi_0$, the dark sector's equation of state can momentarily behave as dark radiation ($\bar{w}_{\rm dark}\approx 1/3$) and dark energy ($\bar{w}_{\rm dark}\approx -1$). These values of $\bar{w}_{\rm dark}$ are well outside of the allowed ranges of equation of state, Eq.~\eqref{eq:wlimits}, which apply for the scale-factor range $a^{-5}-a^{-2.5}$. Hence, parameter space that predicts $\phi$ oscillating around $\phi_0$ during this scale-factor range is almost certainly ruled out. Thus, we expect parameter space that simultaneously satisfies the following three conditions to be ruled out: (1) in the large mass-change case ($\phi_{\rm rise}\rightarrow\phi_0\rightarrow\phi_*\rightarrow\phi_{\rm DM}$) (2) $\phi_{\rm rise}\rightarrow \phi_0$ transition occurs at the latest before the end of bin 4, at a scale factor $a\lesssim 10^{-2.5}$ (3) $\phi_0\rightarrow\phi_*$ occurs at the earliest after the start of bin 1, at a scale factor $a\gtrsim 10^{-5}$. These considerations predict that parameter space points that satisfy the following three inequalities are ruled out
\begin{align}
    &\text{Large mass-change}:\,\,\left(\frac{3\alpha_{\chi\chi}f_\chi}{4}\right)\left(\frac{\lambda_\phi}{H_{\rm eq}^{-1}}\right)^{\frac{1}{2}}\gtrsim 1 ,\label{eq:bglimit1}\\
    &a_{\phi_{\rm rise}\rightarrow\phi_0}\lesssim 10^{-2.5}:\,\,\frac{3\alpha_{\chi\chi}f_\chi}{4}\gtrsim\frac{a_{\rm eq}}{10^{-2.5}},\label{eq:bglimit2}\\
    &a_{\phi_0\rightarrow\phi_*}\gtrsim 10^{-5}:\,\,\left(\frac{3\alpha_{\chi\chi}f_\chi}{4}\right)^{\frac{1}{3}}\left(\frac{\lambda_\phi}{H_{\rm eq}^{-1}}\right)^{\frac{2}{3}}\gtrsim\frac{10^{-5}}{a_{\rm eq}}.\label{eq:bglimit3}
\end{align}
This matches well with the numerically obtained limits shown in Fig.~\ref{fig:results}, except for the beige branch that sticks out. This branch is actually part of a strip corresponding to $(3\alpha_{\chi\chi} f_\chi/4)(\lambda_\phi H_{\rm eq})^{1/2}\sim 1$, that is ruled out by our analysis but is not captured by the approximate conditions of Eqs.~\eqref{eq:bglimit1}, \eqref{eq:bglimit2},\&\eqref{eq:bglimit3}. We discuss this strip in the next subsection.

\subsubsection{Mixed-Dark-Matter Strip and Variation in $\chi$ Abundance}
\label{ss:mixedDM}

As described in Section~\ref{s:background}, the oscillation of $\phi$ around $\phi_*$ eventually behaves like CDM. Therefore, $\phi$ contributes to the DM abundance at late times alongside $\chi$. We can parametrize this contribution from $\phi$ the same way we did with $\chi$ in Eq.~\eqref{eq:fchi}, namely as $f_\phi=\rho_{\phi,0}/\rho_{m,0}$. In nearly all cases, this contribution is negligible compared to that of $\chi$. However, close to the boundary between the regimes of small and large mass excursions, $(3\alpha_{\chi\chi}f_\chi/4) (\lambda_\phi H_{\rm eq})^{1/2}\sim 1$, the contribution from $\phi$ can be comparable to $\chi$'s, $f_\phi/f_\chi=\mathcal{O}(1)$. Hence, we call this approximate line the mixed-DM strip. In Fig.~\ref{fig:results}, a part of the beige region follows this line.

We reiterate that the limits shown in Fig.~\ref{fig:results} were obtained by assuming the fermion $\chi$ makes up the entirety of the DM, which corresponds to setting $f_\chi=f_{\rm DM}=0.85$. Since the mixed-DM strip predicts $f_\phi/f_\chi=\mathcal{O}(1)$, on it the total DM abundance today, as quantified by $f_\chi+f_{\phi}$, would exceed the fiducial value $f_{\rm DM}=0.85$ by $\mathcal{O}(1)$ and would overclose the universe.  Furthermore, there are inherent, ``geometric" degeneracies among cosmological parameters that allow a range of $f_{\rm DM}$ values different from $0.85$ to be consistent with the CMB data \cite{Efstathiou:1998xx,Planck:2018vyg}. Late-time cosmic expansion probes such as SN1a and BAO observations can break these degeneracies to some extent, but still leaves about $20\%$ of $\Omega_{\rm DM,0}$ wiggle room. These uncertainties in $\Omega_{\rm DM,0}$ translate to uncertainties in the allowed value of $f_\chi+f_{\phi}$, and hence $f_\chi$, since the value of $f_\phi$ is set once $f_\chi$ is specified.

The existence of the mixed-DM strip and the uncertainty in the allowed values of $f_\chi$ suggest that we need to reassess the validity of the beige limit in Fig.~\ref{fig:results} which was produced with the assumption that $f_\chi=0.85$, especially along the mixed-DM strip. We do so by checking the robustness of the limits against variations in $f_\chi$. In general, we find that changing $f_\chi$ by $\mathcal{O}(1)$ in the range $0.7-1.2$ amounts to mild (and not necessarily monotonic) shifts in the boundary of the region ruled out by each of the four scale-factor bins. For points deep inside the boundary of at least one of the bins, it is clear that there is no value of $f_\chi$ that yield $f_\chi+f_\phi$ consistent with SN1a and BAO that would save the points from being ruled out. Points along the edges of the ruled out region, including those in the mixed-DM strip, are close to the boundaries of two or more bins, which suggests that they are more likely to be salvageable. However, we checked multiple such points by varying $f_\chi/f_{\rm DM}$ in the range $0.7-1.2$, and found that over this range of $f_\chi/f_{\rm DM}$ they remain ruled out by at least one of the bins, thus demonstrating the robustness of our background-cosmology limits.

\subsection{Limits from Cosmic Perturbations}
Our limits on the dark-force parameter space based on cosmic-perturbation analysis are shown in the right panel of Fig.~\ref{fig:results} in dark blue. These limits gradually worsen with decreasing $\lambda_\phi$, but can in principle extend to very small values of $\lambda_\phi$, as shown in in dotted and dashed dark-blue lines in the left panel of Fig.~\ref{fig:results}. However, at some point these limits are invalidated by model-dependent considerations, i.e., finite-density corrections to the mass and self-interactions of the $\phi$ field, as discussed in Appendix.~\ref{appendix:finitedensity}.

At each parameter-space point, we solve the Meszaros-like Eq.~\eqref{eq:Meszaroslike} for $\delta_\chi$ until the epoch of recombination. We then compute the relative growth factor defined in Eq.~\eqref{eq:relgrowthfactor} and deem the parameter space point ruled out if it satisfies Eq.~\eqref{eq:GrowthThresholds}. The obtained limits are well-approximated by the following conditions: (1) the $\phi\approx\phi_*$ period exists: $2a_{\phi_0=\phi_*}\lesssim \text{min}[a_{\phi_*=\phi_{\rm DM}},a_{\rm rec}]$ and (2) the dark force is effectively stronger than gravity when it is short-range compared to the physical wavenumber $k_{\rm probe}/a$ of the mode of interest ($k_{\rm probe}\lambda_\phi/a\gg 1$): $[(3H^2/2)\Omega_\chi\alpha_{\chi\chi} (k_{\rm probe}\lambda_\phi/a)^2]_{2a_{\phi_0=\phi_*}}\gg 1$. In terms of $\alpha_{\chi\chi}$ and $\lambda_\phi$, this predicts that regimes satisfying following are ruled out
\begin{align}
    \left(\frac{\lambda_\phi}{H_{\rm eq}^{-1}}\right)^{-1/2}\lesssim \alpha_{\chi\chi}&\lesssim \left(\frac{\lambda_\phi}{H_{\rm eq}^{-1}}\right)^{-2}\label{eq:analyticalpertlimit1}\\
    \alpha_{\chi\chi} &\gg \left(\frac{\lambda_\phi}{H_{\rm eq}^{-1}}\right)^{-2}\left(\frac{k_{\rm probe}}{k_{\rm eq}}\right)^{-3}\label{eq:analyticalpertlimit2}
\end{align}
where $k_{\rm eq}\approx 0.01\text{ Mpc}^{-1}$ and we have set $f_\chi=0.85$. These limits provide a good approximation to the numerically obtained ones shown in Fig.~\ref{fig:results}. Note that the limits we obtained are not sensitive to the exact numerical values of the thresholds listed in \eqref{eq:GrowthThresholds}, which we have chosen very conservatively.  Accordingly, we have neglected various $\mathcal{O}(1)$ factors in writing Eqs.~\eqref{eq:analyticalpertlimit1}
and \eqref{eq:analyticalpertlimit2}.

In our analysis, we do not consider the evolution of cosmic perturbations prior to the epoch when the scalar $\phi$ field is tracking its quadratic minimum, $ \phi \approx \phi_*$. It is possible that during this earlier phase, some Fourier modes could grow to nonlinearity and collapse into bound clumps. If so, the typical inter-clump separation could exceed the range of the dark force, in which case dark-force effects between clumps would effectively switch off. Quantifying these effects requires a dedicated nonlinear treatment beyond the linear analysis used here. Nevertheless, it is unlikely that all of the dark sector ends up in compact objects. An $\mathcal{O}(1)$ fraction of the DM  could still remain smooth and continue to participate in linear perturbation dynamics as captured by our analysis. Moreover, nonlinear clumping would imprint additional small-$k$ power, which is readily constrained by the cosmic-perturbation probes discussed here.

\section{Repulsive Dark Forces}
\label{s:repulsive}

\begin{figure}
    \centering
    \includegraphics[width=\linewidth]{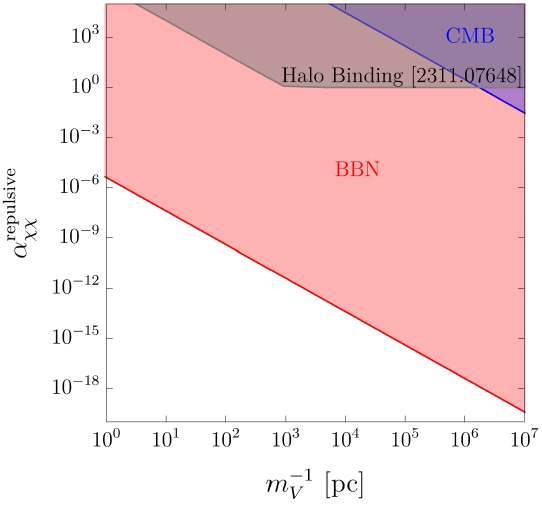}
    \caption{Limits on repulsive dark forces. In the red (blue) region the energy density of the universe is dominated by the scalar potential's contribution $m_V^2(V^0)^2/2\propto z^{-6}$  at redshifts $z\sim 10^{10}$ ($z\sim 10^3$), thus spoiling BBN (CMB). The region labeled ``Halo Binding" is ruled out by the observation of ultrafaint dwarf galaxies, because the repulsive forces would overcome gravity and destabilize the dwarf galaxies \cite{Bogorad:2023wzn}.}
    \label{fig:repulsivelimits}
\end{figure}

The simplest way to make the DM interact repulsively is to give it a net charge under a dark photon. We shall see that the presence of a net charge density automatically leads to an energy component that scales as a stiff fluid $m_A^2(A^0)^2\propto (J^0)^2\propto a^{-6}$, since $J^0\propto n_\chi\propto a^{-3}$. Even if this component makes up a tiny component of the universe today, it would rapidly become increasingly important as we go back in time, potentially spoiling, e.g., BBN's predictions. It can already be seen that cosmology will place strong constraints on such models. We show in this section that these simple considerations yield limits on repulsive dark forces that are even stronger than the limits we have obtained for the attractive Yukawa case.

The Lagrangian that we use to model a repulsive force acting on DM is 
\begin{equation}
    \mathcal{L}\supset -\frac{1}{4}V_{\mu\nu}V^{\mu\nu}+\frac{1}{2}m_V^2V_\mu V^\mu-V_\mu J_{(\chi)}^\mu.
\end{equation}
We assume that the DM $\chi$ is net charged under a dark $U(1)$ gauge boson $V^\mu$ whose mass is $m_V$ and field strength tensor is $V^{\mu\nu}$. We treat $\chi$ as a non-relativistic fluid whose dark-charge current is given by 
\begin{align}
    J_{(\chi)}^\mu=\frac{q_\chi\rho_{\chi}}{m_\chi}u_{(\chi)}^\mu
\end{align}
where $\rho_{\chi}$, $q_\chi$, and $u_{(\chi)}^\mu$ are, respectively, $\chi$'s mass density, $\chi$ particle's dark $U(1)$ charge and $\chi$ fluid's 4-velocity. The Proca equation reads
\begin{align}
    \nabla_\nu V^{\nu\mu}+m_V^2V^\mu=J_{(\chi)}^\mu.
\end{align}
We assume that the universe is homogeneous and isotropic at the background level, and this implies $V^i=0$ and $\nabla_\nu V^{\nu\mu}=0$ (since spatial derivatives must be zero). The Proca equation then tells us that a background dark $U(1)$ charge density $J_{(\chi)}^0$ sources a background scalar potential
\begin{equation}
V^0=J_{(\chi)}^0/m_V^2.
\end{equation}
This highlights an important difference between the vector and scalar cases. In contrast to the heuristically similar ``quasi-static'' solution $\phi\approx \phi_* = gn_\chi/m_\phi^2$ for the scalar which requires ignoring the effects of time-derivatives on the dynamics, the relation $V^0=J_{(\chi)}^0/m_V^2$ in the vector case applies in general for a homogeneous and isotropic universe. In other words, the scalar $\phi$'s ``inertia'' (i.e. second time-derivative term in the equation of motion) prevents it from perfectly tracking its (in general changing) potential minimum. In contrast, the $V^0$ component of a homogeneous vector field background does not have a kinetic term in its equation of motion, and it thereby  ``locks'' onto the instantaneous value of the charge density $J_{(\chi)}^0$.

The scalar potential $V^0$ carries an energy density
\begin{equation}
    \frac{1}{2}m_V^2\left(V^0\right)^2=\frac{1}{2}\left(\frac{J_{(\chi)}^0}{m_V}\right)^2\propto a^{-6}.
\end{equation}
Its scaling as $a^{-6}$ follows from DM particle-number conservation which ensures that the charge density scales as $J_{(\chi)}^0\propto a^{-3}$. Knowing this scaling, we compute the energy density of the background scalar potential $V^0$ at an arbitrary redshift $z$ as
\begin{align}
    \frac{1}{2}m_V^2\left(V^0\right)^2&\sim \frac{q^2\rho_{\chi,0}^2z^6}{m_\chi^2m_V^2}\nonumber\\
    &\sim 3\times 10^5\MeV^4\alpha_{\chi\chi}^{\rm repulsive}\left(\frac{m_V^{-1}}{\text{pc}}\right)^2\left(\frac{z}{10^{10}}\right)^6
\end{align}
where we have defined the coupling strength relative to gravity as
\begin{align}
    \alpha_{\chi\chi}^{\rm repulsive}\equiv \frac{q^2}{4\pi Gm_\chi^2}
\end{align}
and set $\rho_{\chi,0}\approx 11\meV^4$ \cite{Planck:2018vyg}. Requiring $m_V^2(V^0)^2/2$ at redshift $z_{\rm BBN}\sim \text{ MeV}/(0.1\text{ meV})\sim 10^{10}$ be much less than $\text{MeV}^4$ in order to avoid spoiling nucleosynthesis, we find a rough upper bound on $\alpha_{\chi\chi}^{\rm repulsive}$
\begin{align}
    \alpha_{\chi\chi}^{\rm repulsive}\lesssim 4\times 10^{-6}\left(\frac{m_V^{-1}}{\text{pc}}\right)^{-2} &&\text{(BBN)}
\end{align}
which is far more stringent than the limits we found in the main text on attractive Yukawa forces. Furthermore, requiring $m_V^2(V^0)^2/2\lesssim \eV^4$ at $z\sim 10^3$ yields
\begin{align}
    \alpha_{\chi\chi}^{\rm repulsive}\lesssim 3\times 10^{12}\left(\frac{m_V^{-1}}{\text{pc}}\right)^{-2} &&\text{(CMB)}
\end{align}
If the above is not satisfied then the universe would be stiff-fluid dominated at what is supposed to be the epoch of matter-radiation equality. Such a cosmology is inconsistent with CMB data and is therefore ruled out.

The current best astrophysical limit on $\alpha_{\chi\chi}^{\rm repulsive}$ comes from the existence of ultrafaint dwarf galaxies \cite{Bogorad:2023wzn} and is approximately given by $\alpha_{\chi\chi}^{\rm repulsive}\lesssim 1+(m_V^{-1}/\text{kpc})^{-2}$. Dwarf galaxies would not be stable if the strength of repulsive dark forces exceeds this limit. We summarize the crude limits on repulsive dark forces discussed in this section in Fig.~\ref{fig:repulsivelimits}.
We see that the bound from BBN is much stronger than the astrophysical bound and also stronger than the bounds on an attractive force.  We should note that the BBN bound is somewhat model-dependent.  In particular we would not have to satisfy the BBN bound if the dark matter does not exist at the time of BBN and is only created later in the universe.  We do not need to have DM at the time of BBN, so in principle the BBN bound does not need to be satisfied, however one would need to carefully build a DM model specifically to avoid it.  Generic models for such DM would generally be subject to this BBN bound.

\section{Discussion and Conclusion}
\label{s:conclusion}
We have explored the cosmological implications of an attractive long-range force acting between DM particles which is stronger than gravity. We considered a model where the DM is fermionic and the dark force is mediated by a light scalar coupled to the fermionic DM. We found that such a dark sector can behave, at the background level, momentarily as dark radiation and dark energy, in stark deviation from cold dark matter, which is known to simultaneously fit a wide range of cosmological observables very well. At the perturbation level, dark forces can cause orders of magnitude of extra growth in small-scale perturbations, enough to be observable in existing probes of small-scale power spectra, including but not limited to  Lyman-$\alpha$ forest and ultrafaint dwarf heating. Our analysis places some of the strongest constraints on the strength $\alpha_{\chi\chi}$ of dark forces with ranges $\lambda_\phi\lesssim 1\text{ Mpc}$, at least for the Yukawa model under consideration.

The Yukawa model is, of course, not the only possible dark-force model. Nevertheless, it illustrates a range of possible early-universe dynamics one might find in a generic dark-force model, as well as the observable imprints it may leave. As another example, we briefly considered a model of repulsive dark forces, where the DM is net charged under a dark photon. By requiring consistency with BBN, we placed a limit on the strength of repulsive dark forces that is even more stringent than what we have obtained for the attractive case. It would be interesting to consider other dark-force models \cite{Chakraborty:2022mwu,Chakraborty:2024xas}. While we focus on inter-DM forces in this work, the mechanisms we studied here could be relevant for any ultralight scalars coupled to fermions, e.g. neutrinos \cite{Fardon:2003eh,Kaplan:2004dq,Sakstein:2019fmf,Berryman:2022hds,Banerjee:2025nvs}.

The newly obtained limits on long-range inter-DM forces, combined with existing limits on, say, inter-nucleon fifth forces, can be translated into new limits on DM-nucleon forces. In known models of long-range forces, the DM-DM coupling $\alpha_{\chi\chi}$, nucleon-nucleon coupling $\alpha_{nn}$, and DM-nucleon coupling $\alpha_{\chi n}$ typically satisfy the geometric mean relation $\alpha_{\chi n}=\sqrt{\alpha_{\chi\chi}\alpha_{nn}}$. At $\lambda_\phi=1\text{ AU}$, for instance, in the attractive case our perturbation analysis constrains $\alpha_{\chi\chi}\lesssim 10^{8}$. This, in combination with the current limit on inter-nucleon forces from MICROSCOPE, $\alpha_{nn}\lesssim 2\times 10^{-12}$, implies that $\alpha_{\chi n}\lesssim 10^{-2}$ at AU scales. Since the current strongest limits on $\alpha_{\chi n}$ \cite{Bogorad:2024hfj} are obtained with a similar method of taking the geometric mean of the limits on $\alpha_{\chi\chi}$ and on $\alpha_{nn}$, our limits on $\alpha_{\chi\chi}$ also imply the strongest limits on $\alpha_{\chi n}$, for a wide range of $\lambda_\phi$. This will have important implications, as DM-nucleon interactions far stronger than the limit we have obtained were often assumed in astrophysical contexts up till now.

We have seen that in the Yukawa model under consideration, the dark sector goes through phases with equation of state $w_{\rm dark}\approx 1/3$ and $w_{\rm dark}\approx -1$ when $\phi$ is oscillating around the linear-potential minimum $\phi_0$. In certain parameter space, around $\alpha_{\chi\chi}\sim 1$ and $\lambda_\phi\sim H_{\rm eq}^{-1}\approx 30\text{ kpc}$, the $w_{\rm dark}\approx 1/3$ phase is nearly non-existent, while the $w_{\rm dark}\approx -1$ phase lasts a long time and occurs around matter-radiation equality. Since this behavior is similar to the early dark energy (EDE) scenario \cite{Poulin:2018cxd,Kamionkowski:2022pkx}, it might be interesting to revisit this parameter space with a more detailed analysis. Unlike other EDE scenarios, which consider an extra dark component in an otherwise $\Lambda$CDM cosmology, we have a dark energy component that becomes the dark matter. Our scenario is thus qualitatively different from the standard EDE as it changes the evolution of the DM and does not merely add an extra energy component to $\Lambda$CDM.

At the perturbation level, the effects of long-range dark forces are not limited to enhanced matter power on small scales. They may also have important implications for the abundances and properties of DM halos. Dark forces may cause structures to collapse earlier than in the standard scenario, forming so-called ultracompact minihalos. Other considerations that we do not include here, such as NANOGrav's non-detection of scalar-induced GWs and microlensing limits on primordial black holes and other compact structures, may strengthen the limits on dark forces further \cite{Iovino:2024tyg, Inomata:2018epa}. These probes may access yet higher wavenumbers and potentially extend the dark-force limits to even shorter $\lambda_\phi$.

There are possible caveats to our results. We have assumed in our analysis that the primordial power spectrum is scale invariant. While it is possible that the primordial power spectrum has a nontrivial running of scalar spectral index at higher wavenumbers than probed by the CMB, it would require a tuning for the non-standard primordial spectrum to exactly cancel out the scale-dependent growth enhancement due to dark forces. However, effects that lead to the dissipation of high-$k$ perturbation power could serve as potential caveats, as they may wash away any prior enhancement caused by dark-forces. For instance, the $\chi$ particles may have non-negligible temperature by virtue of their production mechanism. In that case, they may cause enough diffusion damping to undo dark-force clustering effects on scales smaller than their free-streaming length. Apart from that, we reiterate that the scalar field $\phi$'s bare mass and bare quartic coupling considered in this work are generally smaller than their expected loop corrections. It would be interesting to address the model-building issues surrounding such fine-tuned scenarios and to consider the effects of larger quartic couplings in the future.

Earlier works prior to ours that study the cosmological implications of dark forces tend to focus on the small coupling $\alpha_{\chi\chi}$ regime, where the effects of the dark force amount to a small correction to the dominant gravitational force. In this regime, it is essential to perform a detailed likelihood analysis, as done in, e.g., \cite{Bottaro:2024pcb}, to disentangle the effects of dark forces from the degeneracies of cosmological parameters. Although these analyses are highly valuable, the practically accessible region of the $(\alpha_{\chi\chi}, \lambda_\phi)$ parameter space is limited to suitably small couplings, where the perturbative expansion in $\alpha_{\chi\chi}$ remains valid, and to $\lambda_\phi \gtrsim H_{\rm eq}^{-1}$, beyond which the computational cost of the analysis becomes prohibitive. Our analysis is complementary to such detailed numerical analyses as it covers the large coupling regime, $\alpha_{\chi\chi}\gtrsim 1$, and is far less computationally limited at smaller $\lambda_\phi$ values. In this regime of the parameter space, the cosmology is in such stark deviation from $\Lambda$CDM that a detailed numerical analysis would not be necessary to rule it out. Thus, our analysis offers a relatively economical way to cover this otherwise numerically and technically challenging regime.

\acknowledgments
We thank Tom Abel, Cyril Creque-Sarbinowski, Sten Delos, Savas Dimopoulos, Dan Green, Junwu Huang, David~E.~Kaplan, Xuheng Luo, Joel Meyers, Oliver Philcox, Surjeet Rajendran, Tristan Smith, Ken Van Tilburg, and Zachary Weiner for useful discussions. This work was supported in part by NSF Grant No.~PHY-2310429, NSF Grant No. PHY-2515007, Simons Investigator Award No.~824870,  the Gordon and Betty Moore Foundation Grant No.~GBMF7946, The
University of Delaware Research Foundation and the John Templeton Foundation Award No.~63595.

\appendix
\section{Scalar Evolution Details}
\label{appendix:scalardetail}
It is useful to define $x\equiv a/a_{\rm eq}$ and write the Klein-Gordon equation and the Hubble rate in terms of $x$
\begin{widetext}
    \begin{align}
        0&=\frac{d^2\phi}{dx^2}+\left(\frac{1}{H}\frac{dH}{dx}+\frac{4}{x}\right)\frac{d\phi}{dx}+\frac{m_\phi^2}{H^2x^2}\left[\phi-\phi_*\text{sign}\left(\phi_0-\phi\right)\right]\label{eq:eomapp}\\
        H^2&=\frac{H_{\rm eq}^2}{2}\left(\frac{1-f_\chi}{x^3}+\frac{1}{x^4}\right)+\frac{H_{\rm eq}^2}{2}\left(\frac{f_\chi|1-\phi/\phi_0|}{x^3}\right)+\frac{\dot{\phi}^2}{2}+\frac{m_\phi^2\phi^2}{2}
    \end{align}
\end{widetext}
where 
\begin{align}
    f_\chi=\frac{\Omega_{\chi,0}}{\Omega_{\rm m,0}}
\end{align}
is the fraction of the present-day matter density in $\chi$. It will be set by the requirement that $\rho_{\rm DM,0}/\rho_{\rm m,0}\approx 0.85$. We start by assuming that the Hubble rate $H$ is the same as in $\Lambda$CDM to the zeroth order. Later, we will check the validity of this assumption \textit{a posteriori}. During the course of its evolution, the envelope of the background $\phi$ can track or oscillate around the following special solutions
\begin{align}
    \phi_{\rm rise}&=\frac{3\alpha_{\chi\chi}f_\chi }{4}\phi_0x\label{eq:phiriseapp}\\
     \phi_0&=\sqrt{\frac{2}{\alpha_{\chi\chi}}}M_{\rm pl}\\
    \phi_*&=\frac{3\alpha_{\chi\chi}f_\chi }{4}\left(\frac{H_{\rm eq}}{m_\phi}\right)^2 \frac{\phi_0}{x^3}\\
    \phi_{\rm DM}^{\rm rms}&= \sqrt{\frac{3\alpha_{\chi\chi}f_\phi}{4}}\left(\frac{H_{\rm eq}}{m_\phi}\right)\frac{\phi_0}{x^{3/2}}
\end{align}
where $M_{\rm pl}\equiv 1/\sqrt{8\pi G}=2.4\times 10^{18}\GeV$ is the reduced Planck mass; $\phi_{\rm rise}$ is an attractor during RD when $\phi\ll \phi_*,\phi_0$; $\phi_0\equiv m_\chi/g$ is the linear finite-density minimum; $\phi_*\equiv gn_\chi/m_\phi^2$ is the quadratic finite-density minimum; $\phi_{\rm DM}$ is the rms amplitude of a dark-matter like solution normalized to the fraction of the total dark matter it comprises today, $f_\phi=\rho_{\phi,0}/\rho_{\rm DM,0}$.

There are two types of evolution histories that the envelope of $\phi$ can take:
\begin{itemize}
    \item \textit{Large mass-change case} ($\phi_{\rm rise}\rightarrow\phi_0\rightarrow\phi_{*}\rightarrow\phi_{\rm DM}$):\\  If $3\alpha_{\chi\chi}f_\chi/4\gtrsim (m_\phi/H_{\rm eq})^{1/2}$, then $\phi$ does reach $\phi_0$ during its evolution, which goes schematically as follows: $\phi_{\rm rise}\rightarrow \phi_0\rightarrow\phi_*\rightarrow\phi_{\rm DM}$. First $\phi$ approaches $\phi_{\rm rise}\propto x$ and proceeds to track it, and then $\phi$ tracks $\phi_0=\text{constant}$ on average but oscillates with a bottom-potential velocity of $|\dot{\phi}_{\rm lin}|\approx H(x)\phi_0\propto x^{-2}$, then $\phi$ tracks $\phi_*\propto x^{-3}$ adiabatically with $\dot{\phi}\approx \dot{\phi}_*\propto x^{-5}$ for a while, before redshifting like matter  with a root-mean-square (rms) field-space velocity $\dot{\phi}_{\rm DM}^{\rm rms}\sim \left|\dot{\phi}_*\right|_{\phi_0=\phi_*}(x/x_{\phi_0=\phi_*})^{-3/2}$. In this case, the points where $\phi$ switches behavior are
    \begin{align}
    x_{\phi_{\rm rise}=\phi_0}&\approx \left(\frac{3\alpha_{\chi\chi}f_\chi}{4}\right)^{-1}\\
     x_{\phi_0=\phi_*}&\approx \left(\frac{3\alpha_{\chi\chi}f_\chi}{4}\right)^{1/3}\left(\frac{H_{\rm eq}}{m_\phi}\right)^{2/3}\\
    x_{\phi_*=\phi_{\rm DM}^{\rm rms}}&\approx \left(\frac{3\alpha_{\chi\chi}f_\chi}{4}\right)^{7/9}\left(\frac{H_{\rm eq}}{m_\phi}\right)^{8/9}
    \end{align}
    \item \textit{Small mass-change case} ($\phi_{\rm rise}\rightarrow\phi_{\rm DM}$):\\ If $3\alpha_{\chi\chi}f_\chi/4\lesssim (m_\phi/H_{\rm eq})^{1/2}$, then $\phi$ does not reach $\phi_0$ during its evolution, which goes schematically as follows: $\phi_{\rm rise}\rightarrow\phi_{\rm DM}$. First $\phi$ approaches $\phi_{\rm rise}\propto x$ and tracks it, without ever reaching $\phi_0=\text{constant}$, and then it crosses $\phi_*$ at
    \begin{align}
         x_{\phi_{\rm rise}=\phi_*}\approx \left(\frac{H_{\rm eq}}{m_\phi}\right)^{1/2}
    \end{align}
    and from then on redshifts like dark matter with an rms field-space velocity $|\dot{\phi}_{\rm DM}^{\rm rms}|\approx \left|\dot{\phi}_*\right|_{\phi_{\rm rise}=\phi_*}(x/x_{\phi_{\rm rise}=\phi_*})^{-3/2}$. 
\end{itemize}
Note that the large (small) mass-change case corresponds to $x_{\phi_{\rm rise}=\phi_0}\lesssim x_{\phi_{\rm rise}=\phi_*}$ ($x_{\phi_{\rm rise}=\phi_0}\gtrsim x_{\phi_{\rm rise}=\phi_*}$). Below, we derive the piecewise analytical formulas mentioned above.

\subsection{$\phi_{\rm rise}$: initial attractor}

\begin{figure}[t!]
    \centering
    \includegraphics[width=\linewidth]{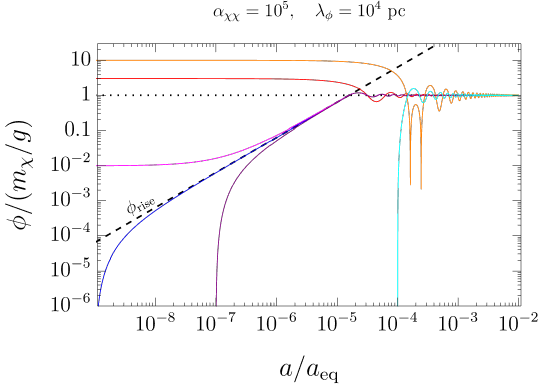}
    \caption{Dynamical attractors of the background evolution of $\phi$. The solid lines are the evolutions of $\phi$ obtained by solving Eq.~\eqref{eq:eomapp} with different initial conditions for $\phi$ (and with $\dot{\phi}=0$ initially). The plot shows that the $\phi_{\rm rise}$ solution defined in Eq.~\eqref{eq:phiriseapp} (dashed line) and also the linear finite-density minimum $\phi_0=m_\chi/g$ (dotted line) are attractors for a wide range of initial conditions. For the assumed values of $\lambda_\phi$ and $\alpha_{\chi\chi}$, the $\phi_*=gn_\chi/m_\phi^2$ is far greater than $\phi_0=m_\chi/g$, and thus lies outside of the range of the plot.}
    \label{fig:attractor}
\end{figure}

Suppose we start deep in RD at $x=x_i\ll 1$ when the Hubble rate is $H(x_i)\approx H_{\rm eq}/\sqrt{2}x_i^2$, the mediator field is $\phi_i$, and its derivative is $\dot{\phi}_i$. It turns out that for a wide range of initial conditions and parameter space, the solution $\phi_{\rm rise}=(3\alpha_{\chi\chi}f_\chi/4)\phi_0x$ is a dynamical attractor for the initial evolution of $\phi$ during RD. When $\phi\ll \phi_*,\phi_0$, the equation of motion for $\phi$ reduces to
\begin{align}
    \frac{d}{dx}\left(x^2\frac{d\phi}{dx}\right)=\frac{3\alpha_{\chi\chi}f_\chi }{2} \phi_0x
\end{align}
whose general solution is
\begin{align}
    \phi=\phi_i+\frac{\dot{\phi}_i}{H(x_i)}\frac{x-x_i}{x}+\frac{3\alpha_{\chi\chi}f_\chi}{4}\phi_0\frac{(x-x_i)^2}{x} \label{eq:attractorsol}
\end{align}
Here, the first two terms are the initial field position and the field excursion due to the initial field velocity, while the last term captures the rolling of the field down the linear, finite-$\chi$-density potential. This last term asymptotes to $\phi_{\rm rise}$ when $x\gg x_{\rm rise}$. Since in this limit $\phi_{\rm rise}\propto x$, it grows relative to the initial-condition terms, and will eventually becomes dominant, at
\begin{align}
    x\sim x_{\rm rise}\equiv \text{max}\left[x_i,\frac{\phi_i+\dot{\phi}_i/H(x_i)}{(3\alpha_{\chi\chi}f_\chi/4)\phi_0}\right]
\end{align}
In other words, $\phi_{\rm rise}$ is an attractor. Fig.~\ref{fig:attractor} illustrates $\phi$'s tendency to approach $\phi_{\rm rise}$ in cases with $\dot{\phi}_i=0$, for different values of $\phi_i$. If $|\phi_i|\ll\phi_{\rm rise}$, then $\phi$ quickly increases toward $\phi_{\rm rise}$ in a few $e$-folds and proceeds to track it. Next, if $\phi_{\rm rise}\ll |\phi_i|<\phi_0$, then $\phi$ remains frozen at $\phi_i$ until $\phi_{\rm rise}$ crosses it, whereupon $\phi$ switches to tracking $\phi_{\rm rise}$. Further, if $|\phi_{i}|>\phi_0$,\footnote{When $|\phi_{i}|>\phi_0$, the $\phi$ field sees $\phi_0$ but not $\phi_*$ as the instantaneous effective-potential minimum; see Fig.~\ref{fig:Veff}.} then $\phi$ would similarly stay frozen at $\phi_i$ until $\phi_{\rm rise}$ crosses it, however this time after the crossing $\phi$ switches to oscillating around $\phi_0$ with an initial-condition-dependent amplitude instead of tracking $\phi_{\rm rise}$. Finally, any initial field velocity $\dot{\phi}_i$ will slow down due to Hubble friction to approach the terminal slow-rolling, which is captured by the third term of Eq.~\eqref{eq:attractorsol}. At $x\gg x_i$, accounting for a nonzero $\dot{\phi}_i$ and amounts to adding a constant field excursion $\approx \dot{\phi}_i/H(x_i)$ to the initial field value of $\phi$.

Overall, $\phi$ would track $\phi_{\rm rise}$ at some point during its evolution if its initial value $\phi_i$ and initial velocity $\dot{\phi}_i$ are sufficiently small. If $\phi$ never tracked $\phi_{\rm rise}$, the late-time evolution of $\phi$ would be initial-condition-dependent and there would be more branching of cases. However, in those cases the dark sector will carry more energy density than in the case where $\phi$ had tracked $\phi_{\rm rise}$, and this translates to larger oscillations around $\phi_0$ and $\phi_*$ at late times. The latter suggests that
deviating from the $\phi_{\rm rise}$-tracking case would lead to background evolutions that are more catastrophic and so easier to rule out. Thus, for the purpose of placing limits on dark forces, it is both generic and conservative to assume the attractor solution $\phi_{\rm rise}$ as an initial condition. We will also assume, for simplicity, that the initial conditions, $\phi_i$ and $\dot{\phi}_i$, are set (by an unspecified earlier dynamics) sufficiently early that the pre-$\phi_{\rm rise}$ evolution does not have observable consequences.

Once $\phi$ reaches $\phi_{\rm rise}$, it will continue tracking $\phi_{\rm rise}$ until whichever of the following comes first: radiation domination ending at $x=1$ or $\phi_{\rm rise}$ crossing the instantaneous minimum of the effective potential of $\phi$, namely either $\phi_0$ or $\phi_*$, whereupon $\phi$ switches into oscillating around $\phi_0$ or $\phi_*$, respectively. If $\lambda_\phi\gtrsim H_{\rm eq}^{-1}$ the $\phi_{\rm rise}$ solution continues into MD, where it asymptotes, at $x\gg 1$, to $\approx \alpha_{\chi\chi} f_\chi\phi_0\ln x$. So, when $\lambda_\phi\gtrsim H_{\rm eq}^{-1}$, $\phi$ never reaches $\phi_0$ if $\alpha_{\chi\chi}f_\chi\ln(10^3)\ll 1$.

\subsection{$\phi_{\rm lin}$: oscillations around $\phi_0$}
The discussion in this subsection is relevant only for the large mass-change case, namely in the parameter space where $\phi_{\rm rise}$ crosses $\phi_0$ before it crosses $\phi_*$.

After the $\phi_{\rm rise}=\phi_0$ crossing point, in the absence of $\phi$ damping other than Hubble, $\phi$ oscillates around $\phi_0$ with a bottom-potential velocity of
\begin{align}
    |\dot{\phi}_{\rm lin}|\approx \left.\dot{\phi}_{\rm rise}\right|_{\phi_{\rm rise=\phi_0}}\approx \phi_0H\propto x^{-2}
\end{align}
Note that the energy density associated to this oscillation scales like radiation, $|\dot{\phi}_{\rm lin}|^2/2\propto x^{-4}$. This kinetic energy becomes negligible compared to the dark energy associated to the mass term $m_\phi^2\phi_0^2/2$ when the Hubble $H$ goes below the mass $m_\phi$. Given a velocity amplitude $\dot{\phi}_{\rm lin}$, the amplitude and period of $\phi-m_\chi/g$ oscillation under the linear potential can be found by simple kinematics
\begin{align}
    \left|\phi_{\rm lin}-\phi_0\right|&=\frac{|\dot{\phi}_{\rm lin}|^2}{2gn_\chi}\approx \frac{\phi_0}{4}\left(\frac{x_{\phi_{\rm rise}=\phi_0}}{x}\right)\\
    T_{\rm lin}&=2\times \frac{2|\dot{\phi}_{\rm lin}|}{gn_\chi}=2\left(\frac{x_{\phi_{\rm rise}=\phi_0}}{x}\right)H^{-1}\propto x
\end{align}
Initially, when $\phi$ first crosses $\phi_0$, a half oscillation takes a Hubble time. Subsequently, each oscillation takes a progressively lesser fraction of the Hubble time, but a longer time to complete.

One might worry that $\chi$ can become relativistic when its effective mass $M_\chi=|m_\chi-g\phi|$ vanishes near $\phi=\phi_0\equiv m_\chi/g$.  The relevant question is how much time $\phi$ spends in the neighborhood of $\phi_0$. The $\chi$ particles are relativistic only where $M_\chi$ drops below their typical momentum $p_\chi$, which is set by the Fermi momentum $p_F$ in the degenerate case (or by the thermal momentum $p_T$ in a thermal production scenario). Because we assume $\chi$ to be extremely cold, $p_\chi$ is tiny, so the relativistic condition $M_\chi\lesssim p_\chi$ is met only within very narrow field-space intervals $\Delta\phi\sim p_\chi/g$ around each $\phi=\phi_0$ crossing. Note that $\left|\phi_{\rm lin}-m_\chi/g\right|$ has the same scaling as that of the momenta of $\chi$ particles, so the typical velocity of $\chi$, $v_\chi=p_\chi/|m_\chi-g\phi|$, does not change with scale factor. In the $T_\chi\approx 0$ case, $\chi$ is non-relativistic  \textit{on average} if
\begin{align}
    \frac{p_F}{g\left|\phi_{\rm lin}-\phi_0\right|}\sim 2\times 10^{-18}\alpha_{\chi\chi}^{1/2}\left(\frac{m_\chi}{\GeV}\right)^{-1/3}\lesssim 1
\end{align}
i.e., $\chi$ is non-relativistic on average for a wide range of $\alpha_{\chi\chi}$ and $m_\chi$. In another appendix section, we show that the $\chi$ particles can be relativistic briefly when $|m_\chi-g\phi|\lesssim p_F\sim n_\chi^{1/3}$. But $\phi$ always swings far away from this region, and outside this region the $\chi$ particles are very non-relativistic. In Eq.~(A18), we find that the fraction of each oscillation spent with $M_\chi\lesssim p_\chi$ is extremely small in the parameter space we consider. Consequently $\chi$ is highly non-relativistic during essentially all of its evolution. Furthermore, non-adiabatic particle production can occur whenever $\phi$ crosses $m_\chi/g$, but this is almost always negligible, and $m_\chi$-dependent; see Appendix~\ref{sss:relativisticregime}.

\subsection{$\phi_{\rm small}$: oscillations around $\phi_*$}

\begin{figure}[t!]
    \centering
    \includegraphics[width=\linewidth]{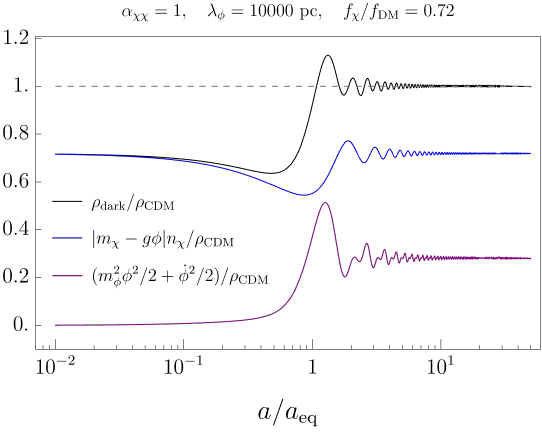}
    \caption{Mixed DM scenario. Here, we plot as function of the scale factor $\chi$'s effective energy density $|m_\chi-g\phi|n_\chi$, $\phi$'s energy density $m_\phi^2\phi^2/2+\dot{\phi}^2/2$, and their sum $\rho_{\rm dark}$, all normalized to the backward-in-time extrapolation of the CDM energy density $\rho_{\rm CDM}$. The plot shows that the initially negligible average energy density of $\phi$ rises considerably at some point and contributes significantly to $\rho_{\rm dark}$ at late times. Thus, the DM at late times is an $\mathcal{O}(1)$ mix of $\chi$ and $\phi$, both behaving as CDM. In this plot, the parameters are chosen to hit the correct total DM abundance today.}
    \label{fig:mixed}
\end{figure}

When the energy density in $\phi$ is sufficiently small, the minimum of $V_{\rm eff}$ is at $\phi_*$. The transition to this quadratic finite-density minimum occurs approximately when $\phi$ crosses $\phi_*$ for the first time. Any relative deviation or velocity of $\phi$ from $\phi_*$ at that time, as a result of its previous dynamics, will contribute to later oscillations of $\phi$ around $\phi_*$, with an amplitude that scales as $x^{-3/2}$. In general, the initial amplitude of this oscillation is set roughly by the relative speed $|\dot{\phi}-\dot{\phi}_*|$ around the moment when the oscillation around $\phi_*$ starts. In the large mass-change case ($\phi_{\rm rise}\rightarrow\phi_0\rightarrow\phi_{*}\rightarrow\phi_{\rm DM}$) this occurs when $\phi_*$ crosses $\phi$ while it is oscillating around $\phi_0$ with a bottom-potential velocity $|\dot{\phi}_{\rm lin}|\approx \phi_0 H$; In the small mass-change case ($\phi_{\rm rise} \rightarrow \phi_{\rm DM}$) this occurs when $\phi_*$ crosses $\phi\approx \phi_{\rm rise}$ that is rising at the rate $\dot{\phi}_{\rm rise}\approx \phi_{\rm rise}H$. Since $\dot{\phi}_*\approx 3H\phi_*$, we have $|\dot{\phi}-\dot{\phi}_*|\sim H\phi_*$ at the respective crossing points, in both cases. Knowing this, it can be shown that the envelope of $\phi$ is given by
\begin{align}
    \phi_{\text{small}}=&\phi_*+\phi_{\rm DM}^{\rm rms}(f_\phi)
\end{align}
where
\begin{align}
    \frac{f_\phi}{f_\chi}\sim
    \begin{cases}
        \displaystyle \frac{9}{4}\left(\frac{3\alpha_{\chi\chi}f_\chi}{4}\right)^{-4/3}\left(\frac{H_{\rm eq}}{m_\phi}\right)^{-2/3}, &\text{large mass-change}\\
        \displaystyle \frac{9}{4}\left(\frac{3\alpha_{\chi\chi}f_\chi}{4}\right) \left(\frac{H_{\rm eq}}{m_\phi}\right)^{1/2}, &\text{small mass-change}
    \end{cases}
\end{align}
This is maximised at the boundary separating the large mass-change and small mass-change cases, namely $(3\alpha_{\chi\chi}f_\chi/4)(H_{\rm eq}/m_\phi)^{1/2}\sim 1$, where $f_\phi/f_\chi\sim 9/4$; see Fig.~\ref{fig:mixed} for the time evolutions of energy densities in this regime. Along that boundary, the $\phi_{\rm DM}$ solution begins when $\phi\sim \phi_0$, which is the largest value $\phi$ can have, modulo oscillations around $\phi_0$.

To fix $f_\chi$, we require
\begin{align}
    f_\chi\left(1+\frac{f_\phi}{f_\chi}\right)=\frac{\Omega_{\rm DM,0}}{\Omega_{\rm m,0}}\approx 0.85
\end{align}

If $x_{\phi_0=\phi_*}\ll x_{\phi_*=\phi_{\rm DM}^{\rm rms}}$, the oscillation of $\phi$ around $\phi_*$ is negligible at first, i.e., $|\phi-\phi_*|\ll \phi_*$. However, since $|\phi-\phi_*|\propto x^{-3/2}$ redshifts more slowly than $\phi_*\propto x^{-3}$, eventually at $x_{\phi_*=\phi_{\rm DM}^{\rm rms}}$ the oscillation around $\phi_*$ takes over as the envelope of $\phi$.

\section{BBN}
\label{appendix:BBN}
The success of Big Bang Nucleosynthesis is not spoiled if at $x_{\rm BBN}\approx 10^{-6}$ we have $\Delta N_{\rm eff}\lesssim 0.5$, which corresponds to $\rho_{\rm dark}/\rho_{\rm crit}\lesssim 1\%$ then. The largest $\rho_{\rm dark}$ achievable for initial conditions that lead to $\phi$ tracking the attractor $\phi_{\rm rise}$ corresponds to the case where $\dot{\phi}\sim \dot{\phi}_{\rm rise}\sim H\phi$ and $\phi\sim \phi_0$, which amounts to $\Omega_{\rm dark}=\rho_{\rm dark}/\rho_{\rm crit}\lesssim (1+m_\phi^2/H^2)/3\alpha_{\chi\chi}$. This maximum density fraction is achieved at BBN if $x_{\phi_{\rm rise}=\phi_0}\sim x_{\rm BBN}$ and $x_{\phi_{\rm rise}=\phi_*}\gtrsim x_{\rm BBN}$, which translate to $(3\alpha_{\chi\chi}f_\chi/4)^{-1}\sim 10^{-6}$ and $(H_{\rm eq}/m_\phi)^{1/2}\gtrsim 10^{-6}$. Thus, for the fiducial $f_\chi=0.85$, we find that the dark density fraction is at most $\rho_{\rm dark}/\rho_{\rm crit}\sim 3\times 10^{-7}$, corresponding to $\alpha_{\chi\chi}\approx 2\times 10^{6}$ and $\lambda_\phi\approx 3\times 10^{-8}\text{ pc}$. Therefore, we conclude that BBN does not place any constraint on the scenarios we consider.

\section{Derivation of Meszaros-Like Perturbation Equation}
\label{appendix:perturbation}
Here, we use mostly positive metric signature, inline with most cosmic-perturbation literature. We work in the conformal Newtonian gauge \cite{Ma:1994dv} and write the metric as
\begin{align}
    ds^2=a^2\left[-\left(1+2\Phi\right)d\tau^2+\left(1-2\Phi\right)dx_i dx^i\right]
\end{align}
where $\tau$ is the conformal time, $\Phi$ is the Newtonian potential, and we have assumed that anisotropic stress is absent. The perturbation parts of the Klein-Gordon equation for $\phi$ and energy-momentum conservation equations for the entire dark sector, $\nabla_\nu T^{\mu\nu}_{\rm dark}=0$, can be written in spatial Fourier space as
\begin{widetext}
\begin{align}
    \delta\phi''+2\mathcal{H}\delta\phi'+\left(a^2m_\phi^2+k^2\right)\delta \phi&= \frac{3\alpha_{\chi\chi}}{2}\mathcal{H}^2\Omega_{\chi}(\delta_\chi+2\Phi)\text{ sign}(\phi_0-\phi)-2a^2m_\phi^2\Phi\phi+4\Phi'\phi'\\
    \delta_{\rm dark}^\prime&=-\left(1+w_{\rm dark}\right)\left(\theta_{\rm dark}-3\Phi^\prime\right)-3\mathcal{H}\left(\frac{\delta p_{\rm dark}}{\delta \rho_{\rm dark}}-w_{\rm dark}\right)\delta_{\rm dark}\\
    \theta_{\rm dark}^\prime &=-\left(1-3w_{\rm dark}\right)\mathcal{H}\theta_{\rm dark}-\frac{w_{\rm dark}^\prime}{1+w_{\rm dark}}\theta_{\rm dark}+\frac{\delta p_{\rm dark}/\delta \rho_{\rm dark}}{1+w_{\rm dark}}k^2\delta_{\rm dark}+k^2\Phi\\
     \delta_{\rm dark}&=\frac{|m_\chi-g\phi|n_\chi\delta_\chi-a^{-2}\phi^{\prime 2}\Phi+a^{-2}\phi^\prime\delta\phi^\prime+ m_\phi^2 \phi \delta \phi}{|m_\chi-g\phi|n_\chi+\frac{1}{2}a^{-2}\phi^{\prime 2}+\frac{1}{2}m_\phi^2\phi^2}\\
    w_{\rm dark}&=\frac{\frac{1}{2}a^{-2}\phi^{\prime 2}-\frac{1}{2}m_\phi^2\phi^2}{|m_\chi-g\phi|n_\chi+\frac{1}{2}a^{-2}\phi^{\prime 2}+\frac{1}{2}m_\phi^2\phi^2}\\
    \delta\rho_{\rm dark}&=  |m_\chi-g\phi|n_\chi\delta_\chi-a^{-2}\phi^{\prime 2}\Phi+a^{-2}\phi^\prime\delta\phi^\prime+ m_\phi^2 \phi \delta \phi\\
    \delta p_{\rm dark}&=-a^{-2}\phi^{\prime 2}\Phi+a^{-2}\phi^\prime\delta\phi^\prime-m_\phi^2 \phi \delta \phi
\end{align}
\end{widetext}
where a prime indicates a derivative with respect to the conformal time $\tau$, $w_{\rm dark}\equiv p_{\rm dark}/\rho_{\rm dark}$, $\delta_{\rm dark}\equiv \delta \rho_{\rm dark}/\rho_{\rm dark}$, $\delta_\chi\equiv \delta n_\chi/n_\chi$, $(\ldots)^\prime=d(\ldots)/d\tau$, and $\theta_{\rm dark}=ik_jv_{\rm dark}^j$ is the velocity divergence of the dark sector fluid.

Assuming $m_\chi\gg g\phi$ and ignoring $\phi'$, $\phi''$, $\delta\phi'$, $\delta\phi''$, and $\Phi$ terms in the equations of motion of $\phi$ and $\delta\phi$, we find 
\begin{align}
    \phi&\approx \phi_*=\frac{gn_\chi}{m_\phi^2}\\
    \delta\phi&\approx \frac{a^2gn_\chi\delta_\chi}{a^2m_\phi^2+k^2}
\end{align}
which, given the $m_\chi\gg g\phi$ assumption, imply $m_\chi n_\chi\gg m_\phi^2\phi^2$,  $m_\chi n_\chi|\delta_\chi|\gg gn_\chi|\delta\phi|$, and
\begin{align}
    w_{\rm dark}&\sim \frac{\phi_*}{\phi_0 }\\
    w_{\rm dark}^\prime &\sim H\frac{\phi_*}{\phi_0}\\
    \frac{\delta p_{\rm dark}}{\delta\rho_{\rm dark}}&\sim \frac{\phi_*}{\phi_0} \label{eq:cssqphistar}\\
    \delta_{\rm dark}&\approx \delta_\chi \label{eq:deltadarkphistar}
\end{align}
Thus, if $\phi_*\ll \phi_0$ (which is implied by the $m_\chi\gg g\phi$ assumption), then the dark sector behaves like matter to leading order: $w_{\rm dark}\ll 1$, $w'_{\rm dark}\ll \mathcal{H}$, and $\delta p_{\rm dark}/\delta \rho_{\rm dark}\ll 1$. The fluid perturbation equations then reduce to
\begin{align}
    \delta_{\rm dark}^\prime&\approx -\theta_{\rm dark}+3\Phi^\prime\\
    \theta_{\rm dark}^\prime&\approx -\mathcal{H}\theta_{\rm dark}+k^2\Phi+k^2\frac{\delta p_{\rm dark}}{\rho_{\rm dark}}
\end{align}
Even though $\delta p_{\rm dark}$ is small, we keep the last term of the second equation because it can be important when $k\gg \mathcal{H}$.  Taking the conformal time derivative of $\delta'_{\rm dark}$ and substituting $\theta'_{\rm dark}$ equation into it, we find
\begin{align}
    \delta_{\rm dark}^{\prime\prime}+\mathcal{H}\delta_{\rm dark}^\prime=3\left(\Phi^{\prime\prime}+\mathcal{H}\Phi^\prime\right)-k^2\Phi-k^2\frac{\delta p_{\rm dark}}{\rho_{\rm dark}}\label{eq:almostMeszaros}
\end{align}
The metric perturbation $\Phi$ is given by the Poisson equation
\begin{align}
    -k^2\Phi&=4\pi G\left(\rho_{\rm rad}\delta_{\rm rad}+\rho_{B}\delta_B+\rho_{\rm dark}\delta_{\rm dark}\right)\nonumber\\
    &\approx 4\pi G\rho_{\rm dark}\delta_{\rm dark}
\end{align}
In the last line, we neglect the radiation and baryon terms. While the baryon and photon fluids are tightly coupled, these contributions oscillate rapidly at subhorizon scales due to radiation pressure and so mostly average out. Then, using Eqs.~\eqref{eq:cssqphistar}\&\eqref{eq:deltadarkphistar}, we simplify Eq.~\eqref{eq:almostMeszaros} to the following Meszaros-like form:
\begin{align}
    \delta_{\chi}^{\prime\prime}+\mathcal{H}\delta_{\chi}^\prime=4\pi Ga^2m_\chi n_\chi\delta_{\chi}\left[\frac{\alpha_{\chi\chi}k^2}{a^2m_\phi^2+k^2}+1\right]
\end{align}

\section{Complete Finite-Density Potential of $\phi$}
\label{appendix:finitedensity}
A finite density, the fermion $\chi$ affects the dynamics of $\phi$ through the term $g\left<\bar{\chi}\chi\right>$ on the RHS of the Klein-Gordon equation. In the main text, we approximate this term as $g\left<\bar{\chi}\chi\right>\approx g n_\chi\text{sign}(m_\chi-g\phi)$. Here, we consider the exact expression of this term:
\begin{align}
    g\left<\bar{\chi}\chi\right>=g\int \frac{d^3p}{(2\pi)^3}\frac{m_\chi-g\phi}{\sqrt{(m_\chi-g\phi)^2+p^2}}f_\chi(p) \label{eq:bilinear}
\end{align}
We will clarify the regime of validity of the aforementioned approximation and briefly discuss possible consequences of going beyond this regime.

First, we derive Eq.~\eqref{eq:bilinear}. We expand the $\chi$ field in creation and annihilation operators,
\begin{align}
    \chi=&\int d\Pi\sum_s\left[a_{s}(\mathbf{p})u_{s}(\mathbf{p})e^{-ip.x}+b_{s}^\dagger(\mathbf{p})v_{s}(\mathbf{p})e^{+ip.x}\right]\\
    \bar{\chi}=&\int d\Pi\sum_s\left[a_{s}^\dagger(\mathbf{p})\bar{u}_{s}(\mathbf{p})e^{+ip.x}+b_{s}(\mathbf{p})\bar{v}_{s}(\mathbf{p})e^{-ip.x}\right]
\end{align}
with $d\Pi=d^3p/[(2\pi)^3\sqrt{2E_{\mathbf{p}}}]$  and $E_{\mathbf{p}}=\sqrt{(m_\chi-g\phi)^2+|\mathbf{p}|^2}$. Multiplying the above, forming $\bar{\chi}\chi$, taking the expectation value, and then using the canonical anti-commutators $\left\{a_r(\mathbf{p}),a_s^\dagger(\mathbf{q})\right\}=(2\pi)^3\delta_{rs}\delta^{(3)}(\mathbf{p}-\mathbf{q})=\left\{b_r(\mathbf{p}),b_s^\dagger(\mathbf{q})\right\}$ (and zero for others) as well as the spinor normalizations $\bar{u}_r(\mathbf{p})u_s(\mathbf{p})=2(m_\chi-g\phi)\delta_{rs}=-\bar{v}_r(\mathbf{p})v_s(\mathbf{p})$ yields
\begin{align}
    \left<\bar{\chi}\chi\right>_{\rm tot}=&\int \frac{d^3p}{(2\pi)^3}\frac{m_\chi-g\phi}{E_{\mathbf{p}}}\nonumber\\
    &\sum_s\left[\left<a_s(\mathbf{p})a_s^\dagger(\mathbf{p})\right>-\left<b_s^\dagger(\mathbf{p})b_s(\mathbf{p})\right>\right]
\end{align}
Inserting
$\left<a_r^\dagger(\mathbf{p})a_s(\mathbf{q})\right>=(2\pi)^3\delta_{rs}\delta^{(3)}(\mathbf{p}-\mathbf{q})f_\chi^{(s)}(p)$ and $\left<b_r(\mathbf{p})b_s^\dagger(\mathbf{q})\right>=(2\pi)^3\delta_{rs}\delta^{(3)}(\mathbf{p}-\mathbf{q})[1-f_{\bar{\chi}}^{(s)}(p)]$, where the $b$ relation uses one of the aforementioned anti-commutation relations, 
\begin{align}
    \left<\bar{\chi}\chi\right>_{\rm tot}=\int \frac{d^3p}{(2\pi)^3}\frac{m_\chi-g\phi}{E_{\mathbf{p}}}\sum_s\left[f_\chi^{(s)}(p)+f_{\bar{\chi}}^{(s)}(p))-1\right]
\end{align}
The ``-1" term is the (divergent) vacuum contribution. Subtracting it,
\begin{align}
    \left<\bar{\chi}\chi\right>&=\left<\bar{\chi}\chi\right>_{\rm tot}-\left<\bar{\chi}\chi\right>_{\rm vac}\nonumber\\
    &=\int \frac{d^3p}{(2\pi)^3}\frac{m_\chi-g\phi}{E_{\mathbf{p}}}\left[f_\chi(p)+f_{\bar{\chi}}(p)\right]
\end{align}
where we have defined $f_{\chi,\bar{\chi}}(p)\equiv \sum_s f_{\chi,\bar{\chi}}^{(s)}(p)$.
In the asymmetric case assumed in main text, $f_{\bar{\chi}}=0$, recovering the expression in Eq.~\eqref{eq:bilinear}.

\subsection{Degenerate Fermions}
Assuming the fermion $\chi$ is degenerate and asymmetrically populated (with no antiparticles), such that $f_\chi(p)\approx \Theta(p_F-p)$, the above integral can be evaluated exactly, giving
\begin{align}
    g\left<\bar{\chi}\chi\right>=&gn_\chi\text{sign}\left(m_\chi-g\phi\right)\left[\frac{M_\chi(\phi)\sqrt{M_\chi^2(\phi)+p_F^2}}{p_F^2}\right.\nonumber\\
     &+\left.\frac{M_\chi^3(\phi)}{p_F^3}\ln\left(\frac{-p_F+\sqrt{M_\chi^2(\phi)+p_F^2}}{M_\chi(\phi)}\right)\right]
\end{align}
where $M_\chi(\phi)=|m_\chi-g\phi|$ is the effective mass of $\chi$ and $p_F=(6\pi^2n_\chi)^{1/3}$ is the Fermi momentum of $\chi$, which is numerically given by
\begin{align}
    p_F&=3\meV\left(\frac{m_\chi}{\GeV}\right)^{-1/3}x^{-1}
\end{align}
Next, we discuss separately the non-relativistic-fermion ($p_F\ll M_\chi(\phi)$ and relativistic-fermion ($p_F\gtrsim M_\chi(\phi)$) regimes.

The temperature of $\chi$ is not important as long as it is much colder than the Fermi energy $p_F^2/2M_\chi$.

\subsubsection{Non-relativistic regime}

Expanding the exact expression of $g\left<\bar{\chi}\chi\right>$ in both $p_F$ and $g\phi$, regardless of the order expansion ($p_F$ first or $g\phi$ first), we find that the leading-order corrections to the effective potential of $\phi$, which can be written as $\delta V(\phi)=\delta m_\phi^2\phi^2/2+\delta \mu\phi^3/3!+\delta \kappa\phi^4/4!$, are
\begin{align}
    |\delta m_\phi^2|\sim \frac{g^2n_\chi p_F^2}{m_\chi^3},\, |\delta \mu|\sim \frac{g^3n_\chi p_F^2}{m_\chi^4},\, |\delta \kappa|\sim \frac{g^4n_\chi p_F^2}{m_\chi^5} \label{eq:finitedensitycorrections}
\end{align}

Typically, the $\propto g n_\chi$ linear-potential term used in the main text far dominates the effective potential of $\phi$. In those cases, the finite-density mass correction is not important even if it were large compared to the bare mass (which is itself negligible compared to the linear potential). However, near the finite-density minimum, $\phi_*=gn_\chi/m_\phi^2$, this linear-potential term is canceled by the bare-mass term $m_\phi^2\phi\approx m_\phi^2\phi_*= gn_\chi$. Thus, the oscillation of $\phi$ around $\phi_*$ depends on the dominant mass term (term $\propto \phi$ in the equation of motion and $\propto \phi^2$ in the effective
potential). During that evolution phase, the mass correction is unimportant if 
\begin{align}
    \frac{|\delta m_\phi^2|}{m_\phi^2}\sim \frac{\phi_*}{\phi_0}\left(\frac{p_F}{m_\chi}\right)^2 \lesssim 1
\end{align}
The earliest time when the finite-density mass-squared correction $\delta m_\phi^2$ might be important is when $\phi$ crosses $\phi_*$ for the first time, namely at $x=x_{\phi_0=\phi_*}$ in the large mass-change case and at $x=x_{\phi_{\rm rise}=\phi_*}$ in the small mass-change case, where $\phi_*/\phi_0\sim 1$ and $\phi_*/\phi_0\sim (3\alpha_{\chi\chi}/4)(H_{\rm eq}/m_\phi)^{1/2}$, respectively.  Unless $m_\chi$ is extremely light (close to the Tremaine-Gunn lower bound of $m_\chi=100\,\eV$), the parameter space where the $\delta m_\phi^2$ is important lies in regimes with very small $\lambda_\phi$, where our limits do not apply.

Anharmonic terms, in particular the cubic term $\delta\mu\phi^3/3$, whose coefficient $\delta\mu$ is lower order in $g$, may excite modes of $\phi$ perturbation around the $\phi$ background, $\delta\phi(x)\equiv\phi(x)-\bar{\phi}(t)$, through narrow parametric resonance. Assuming $\delta m_\phi^2\ll m_\phi^2$, the equation of motion for the spatially Fourier-transformed mode $\delta \phi_k$ reads
\begin{align}
    \delta \ddot{\phi}_k+3H\delta\dot{\phi}_k+\left[\frac{k^2}{a^2}+m_\phi^2+\delta\mu \bar{\phi}(t)\right]\delta\phi_k=0
\end{align}
Mapping this into the Mathieu equation, we find that the first resonance band exists and has the least stringent instability condition. That said, the first band only populates non-relativistic modes which would rejoin the background after a few $e$-folds of redshifting. Thus, we focus on the second band, which populates relativistic modes with $k_{\rm res}\sim m_\phi$, $\delta k_{\rm res}\sim m_\phi\sqrt{q}$, $\Gamma_{\rm Floquet}^{(2)}\sim m_\phi q^2$, $\delta t_{\text{res}}^{(2)}\sim \sqrt{q}H^{-1}$, and $q\sim \delta\mu\bar{\phi}/m_\phi^2$. The earliest time when $\bar{\phi}$ behaves sinusoidally is, again, at $x=x_{\phi_0=\phi_*}$ in the large mass-change case and at $x=x_{\phi_{\rm rise}=\phi_*}$ in the small mass-change case. At these points the amplitude of the sinusoidal part of $\phi$ is $\bar{\phi}\sim \phi_* H/m_\phi$. Requiring $\Gamma^{(2)}_{\rm Floquet}\delta t_{\rm res}^{(2)}\lesssim 10$ then amounts to
\begin{align}
    \left(\frac{\phi_*}{\phi_0}\right)^{5}\left(\frac{H}{m_\phi}\right)^{3/2}\left(\frac{p_F}{m_\chi}\right)^{5}\lesssim 10
\end{align}
Repeating the same arguments for quartic coupling, in which case $q\sim \delta \kappa\bar{\phi}^2/m_\phi^2$, we find
\begin{align}
    \left(\frac{\phi_*}{\phi_0}\right)^{6/5}\left(\frac{m_\phi}{H}\right)\left(\frac{p_F}{m_\chi}\right)^{4/5}\lesssim 10 
\end{align}

The tree-level cubic coefficient $\mu$ and quartic coefficient $\kappa$ are negligible if they are smaller than the values of the finite-density corrections, $\delta \mu$ and $\delta \kappa$, at the earliest point where $\phi$ oscillates sinusoidally, namely at $x=x_{\phi_0=\phi_*}$ in the large mass-change case and at $x=x_{\phi_{\rm rise}=\phi_*}$ in small mass-change case. 
\begin{align}
    \frac{\mu}{m_\phi}&\lesssim\left(\frac{m_\phi}{\phi_*}\right)\left(\frac{m_\phi}{H}\right)^{3/5}\\
    \kappa &\lesssim \left(\frac{m_\phi}{\phi_*}\right)^2\left(\frac{m_\phi}{H}\right)^{8/5}\label{eq:quarticPR}
\end{align}

\begin{figure*}
    \centering
    \includegraphics[width=0.49\linewidth]{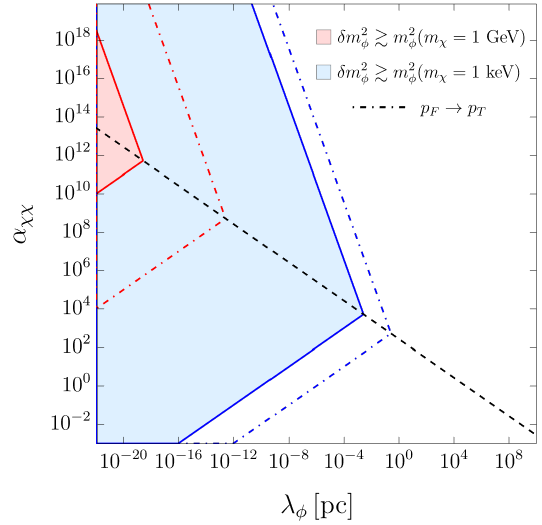}
    \includegraphics[width=0.49\linewidth]{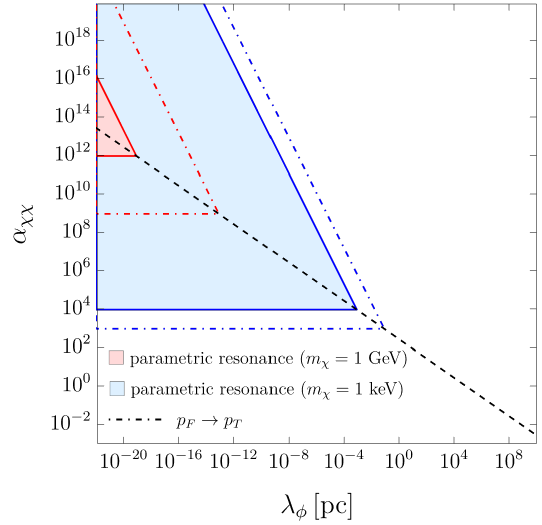}
    \caption{Regimes where finite $\chi$ density and finite $\chi$ temperature are important. \textit{Left: }regimes where the finite $\chi$ density induced mass is important when $\phi$ first oscillates around $\phi_*$. \textit{Right:} regimes where the finite-density induced cubic coupling causes $\phi$ to undergo parametric resonance when $\phi$ starts oscillating around $\phi_*$. In both plots, the shared regions correspond to degenerate $\chi$ particles. The dot dashed lines are where the boundaries would be if $\chi$ has a finite-temperature such that $p_T/p_F$ is given by Eq.~\eqref{eq:pTpF}.}
    \label{fig:placeholder}
\end{figure*}

We display in Fig. 2 parameter space regions where finite-density mass correction and parametric resonance are important for representative values of $m_\chi$.

\subsubsection{Relativistic regime}\label{sss:relativisticregime}
In the limit $M_\chi(\phi)\ll p_F$, which amounts to expanding around $\phi=\phi_0$, we have
\begin{align}
    g\left<\bar{\chi}\chi\right>=\frac{g^2p_F^2}{4\pi^2}(\phi_0-\phi)
\end{align}
Thus, the effective potential of $\phi$ close to $\phi_0$ is quadratic instead of linear, with an effective mass-squared of $m_\phi^2+g^2p_F^2/4\pi^2$. The ratio between the $\chi$-induced mass and the bare mass of $\phi$ is
\begin{align}
    \sim \frac{g n_\chi^{1/3}}{m_\phi}
    &=3\times 10^{4}\alpha_{\chi\chi}^{1/2}\left(\frac{m_\chi}{\GeV}\right)^{2/3}\left(\frac{\lambda_\phi}{1\text{ kpc}}\right)x^{-1}
\end{align}
So, likely very significant, although this is relevant only during the brief moments when $M_\chi(\phi)\ll p_F$.

Every time $\phi$ crosses $\phi_0$, there can be efficient non-adiabatic production of relativistic $\chi$-$\bar{\chi}$ pairs (this is because it is easy to have $|\dot{M}_{\chi}|\gtrsim M_{\chi}^2$ when $M_{\chi}\rightarrow 0$), with typical number density and momentum of
\begin{align}
    n_\chi^{\rm NA}\sim  \frac{|\dot{M}_\chi|^{3/2}}{8\pi^3},\quad 
    p_\chi^{\rm NA}\sim  \frac{|\dot{M}_\chi|^{1/2}}{\pi^{1/2}}
\end{align}
Since $\chi$ is not a boson, the $\chi$ production rate does not grow exponentially. The fractional rate per Hubble time, $\dot{\rho}^{\rm NA}H^{-1}/(\dot{\phi}^2/2)$, at which energy is being drained from $\dot{\phi}^2/2$ to create $\chi$-$\bar{\chi}$ pairs can thus be estimated as
\begin{align}
    \frac{1}{H}\frac{2 p_\chi^{\rm NA}n_\chi^{\rm NA}/T_{\rm lin}}{\dot{\phi}^2/2}\sim \frac{\alpha_{\chi\chi}Gm_\chi^2}{\pi^{5/2}}\frac{x}{x_{\phi_{\rm rise}=\phi_0}}
\end{align}
Since $x_{\phi_{\rm rise}=\phi_0}\sim\alpha_{\chi\chi}^{-1}$. This is negligible unless $m_\chi$ is close to the Planck scale.

\subsection{Non-Degenerate Fermions}

The fermions under consideration would in general have a finite chemical potential but they need not be degenerate. If they were once in thermal equilibrium their occupation number is given by the Fermi-Dirac distribution $f_\chi(p)=(e^{(\sqrt{M_\chi^2+p^2}-\mu)/T_\chi}+1)^{-1}$. Assuming that the fermion is non-relativistic, $p^2/2M_\chi\lesssim T_\chi\ll M_\chi$, we can expand the Fermi-Dirac distribution in $p/M_\chi$, giving $f_\chi(p)\approx (e^{(M_\chi-\mu)/T_\chi}e^{p^2/2M_\chi T_\chi}+1)^{-1}$. In this limit, the chemical potential $\mu$ is related to the total number density of $\chi$ as $n_\chi=(M_\chi T_\chi/2\pi)^{3/2}e^{-(M_\chi-\mu)/T_\chi}$. Thus, we find that $f_{\chi}\sim n_\chi/(p_T/4\pi)^{3/2}$ for $p\lesssim p_T\equiv \sqrt{2M_\chi T_\chi}$ and Boltzmann suppressed for $p\gg p_T$. For crude estimates, we can approximate this distribution function as a step function
\begin{align}
    f_\chi(p)=\frac{2\pi^2n_\chi}{p_T^3}\Theta\left(p_T-p\right)
\end{align}
Since the scalar background is homogeneous, by translational symmetry we can deduce that it will keep momenta of the fermions $\chi$ unchanged. Thus as $M_\chi(\phi)$ is changing, the effective temperature $T_\chi(\phi)=p_T^2/2M_\chi(\phi)$ would change to keep $p_T$ constant. This yields
\begin{align}
    g\left<\bar{\chi}\chi\right>=&gn_\chi\text{sign}\left(m_\chi-g\phi\right)\left[\frac{M_\chi(\phi)\sqrt{M_\chi^2(\phi)+p_T^2}}{p_T^2}\right.\nonumber\\
     &+\left.\frac{M_\chi^3(\phi)}{p_T^3(\phi)}\ln\left(\frac{-p_T+\sqrt{M_\chi^2(\phi)+p_T^2}}{M_\chi(\phi)}\right)\right]
\end{align}
which is similar to the degenerate case. The only difference is that $p_F$ is replaced with $p_T$. The condition for $\delta m_\phi^2\lesssim m_\phi^2$ and $\Gamma_{\rm Floquet}^{(2)}\delta t_{\rm res}^{(2)}\lesssim 10$ thus becomes
\begin{align}
    \frac{\phi_*}{\phi_0}\left(\frac{p_T}{m_\chi}\right)^2&\lesssim 1\\
    \left(\frac{\phi_*}{\phi_0}\right)^{5}\left(\frac{H}{m_\phi}\right)^{3/2}\left(\frac{p_T}{m_\chi}\right)^{5}&\lesssim 10
\end{align}

For instance, if the $\chi$ abundance was set by freeze-out, its typical momentum post freeze-out is given by $p_T\sim \sqrt{2M_\chi T_{\rm FO}}(T/T_{\rm FO})$, where usually $T_{\rm FO}\approx m_\chi/20$.  Since, the Fermi momentum $p_F$ scales the same way, in freeze-out scenarios, $\chi$ can only be degenerate at late times if it is degenerate at freeze out.  In this freeze-out case, the ratio $p_T/p_F$ is approximately
\begin{align}
    \frac{p_T}{p_F}\sim 1\times 10^3\left(\frac{m_\chi}{\GeV}\right)^{1/3} \label{eq:pTpF}
\end{align}
where we have assumed $M_\chi\approx m_\chi$ at freeze out.

If $\chi$ was relativistic in the past, then $g\left<\bar{\chi}\chi\right>$ evaluates to $gM_\chi(\phi) T_\chi^2/3$. If number-changing processes were efficient, $\phi$ could be subject to significant thermal damping \cite{Tanin:2017bzm,Banerjee:2025nvs,Banerjee:2025qyg}.

\section{Naturalness}
\label{Appendix:Naturalness}
\begin{figure}
    \centering
    \includegraphics[width=\linewidth]{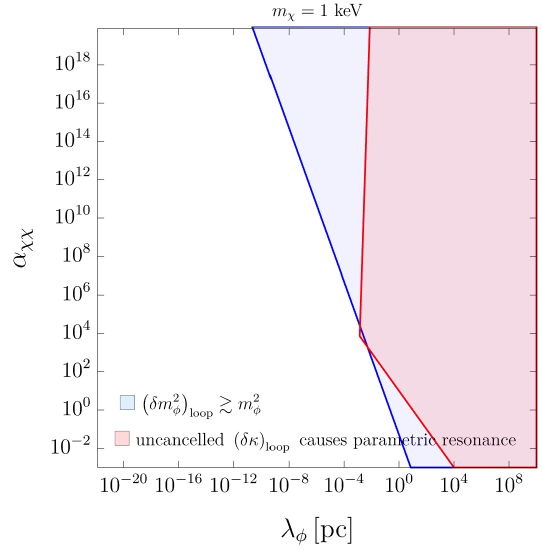}
    \caption{Fine-tuned parameter space for $m_\chi=1\text{ keV}$. In the blue region, the squared mass of the scalar mediator $m_\phi^2$ is smaller than the expected correction to it from $\chi$ loop, $(\delta m_\phi^2)_{\rm loop}$. In the red region, the $\chi$-loop-induced quartic coupling $(\delta\kappa)_{\rm loop}$ would cause parametric resonance, thus invalidating our analysis, unless this $(\delta\kappa)_{\rm loop}$ is somehow canceled by other contributions to the quartic coupling.}
    \label{fig:finetunedparameterspace}
\end{figure}

The bare mass of $\phi$ is greater than the loop correction from $\chi$, $m_\phi^2\gtrsim \left(\delta m_\phi^2\right)_{\rm loop}\sim g^2m_\chi^2/16\pi^2$, if
\begin{align}
    \alpha_{\chi\chi}&\lesssim \frac{32\pi^2M_{\rm pl}^2}{m_\chi^4\lambda_\phi^2}\sim 5\times 10^{-2}\left(\frac{\lambda_\phi}{1\text{ pc}}\right)^{-2}\left(\frac{m_\chi}{\keV}\right)^{-4}
\end{align}
In general, this is difficult to satisfy in the parameter space where our constraints apply. So, the scenarios we consider in this paper are somewhat fine-tuned. For $m_\chi$ values near the Tremaine-Gunn limit, $m_\chi\sim 100\eV$, part of the constrained parameter space can be natural in the above sense, although in the same regime finite-density corrections to $m_\phi^2$ are important, potentially removing the constraints in the first place. The fermion-loop contribution to the quartic coupling $\delta \kappa \phi^4/4$ is roughly $\left(\delta \kappa\right)_{\rm loop}\sim g^4/2\pi^2\sim 4\times 10^{-100}\alpha_{\chi\chi}^2(m_\chi/\keV)^4$. This quartic coupling is negligible if it satisfies Eq.~\eqref{eq:quarticPR}
\begin{align}
    4\times 10^{-100}\alpha_{\chi\chi}^2\left(\frac{m_\chi}{\keV}\right)^4\lesssim \left(\frac{m_\phi}{\phi_*}\right)^2\left(\frac{m_\phi}{H}\right)^{8/5}
\end{align}
We find that in all parameter space constrained by our main analysis, the mediator's mass and quartic coupling are tuned; see Fig.~\ref{fig:finetunedparameterspace}.

\section{Extra Figures}
\label{appendix:extrafigures}

\begin{figure}
    \centering
    \includegraphics[width=\linewidth]{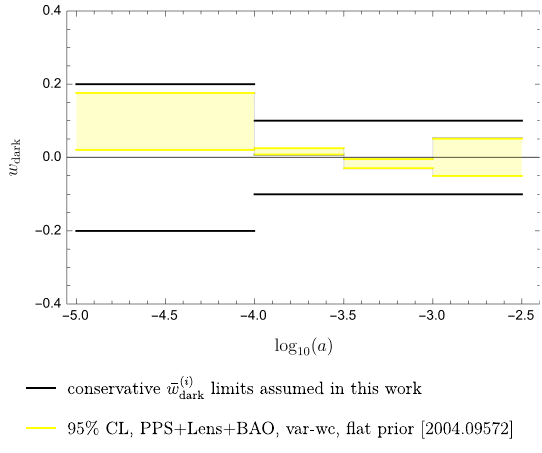}
    \caption{Comparison between the conservative limits on $\bar{w}_{\rm dark}^{(i)}$ we assumed in our analysis, as given in Eq.~\eqref{eq:wlimits}, (black lines) and the allowed values of $\bar{w}_{\rm dark}^{(i)}$ from \cite{Ilic:2020onu} (yellow).}
    \label{fig:wlimitcomparison}
\end{figure}

\begin{figure}
    \centering
    \includegraphics[width=\linewidth]{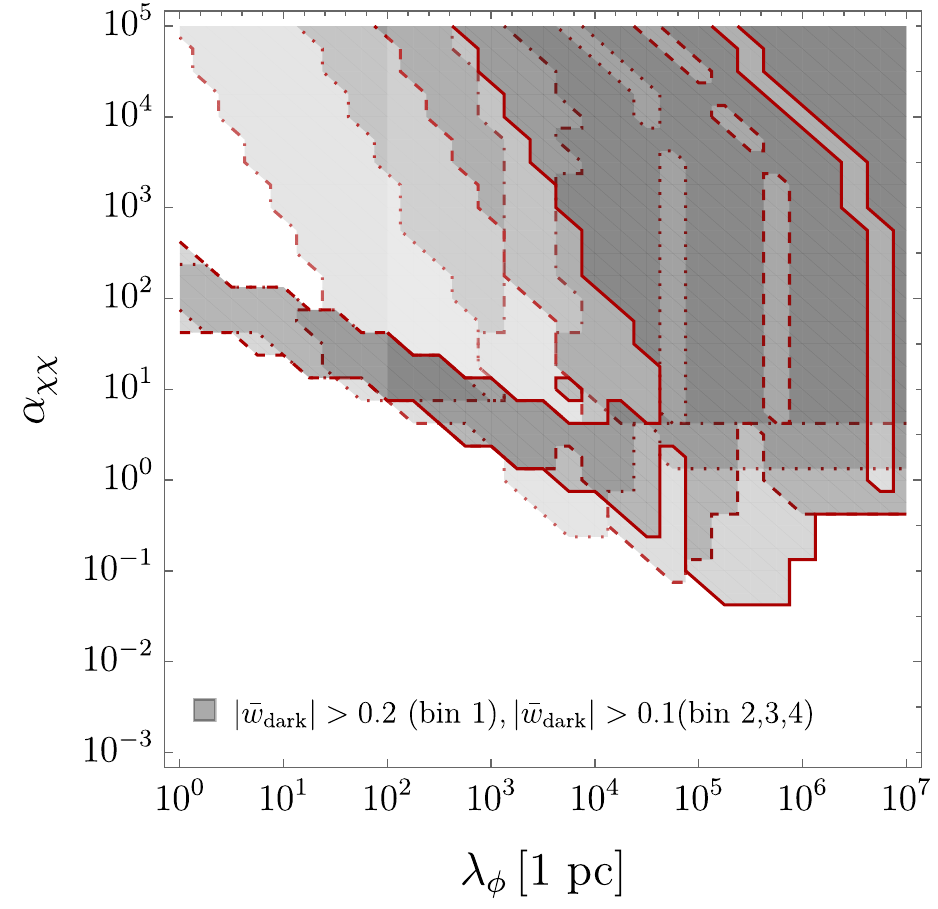}
    \caption{Background limits from each scale-factor bin, Eq.~\eqref{eq:abins}. The limits from bin 1, 2, 3, and 4 are shown in solid, dashed, dotted, and dotdashed.}
    \label{fig:wlimitsperbin}
\end{figure}

\begin{figure*}
    \centering
    \includegraphics[width=0.495\linewidth]{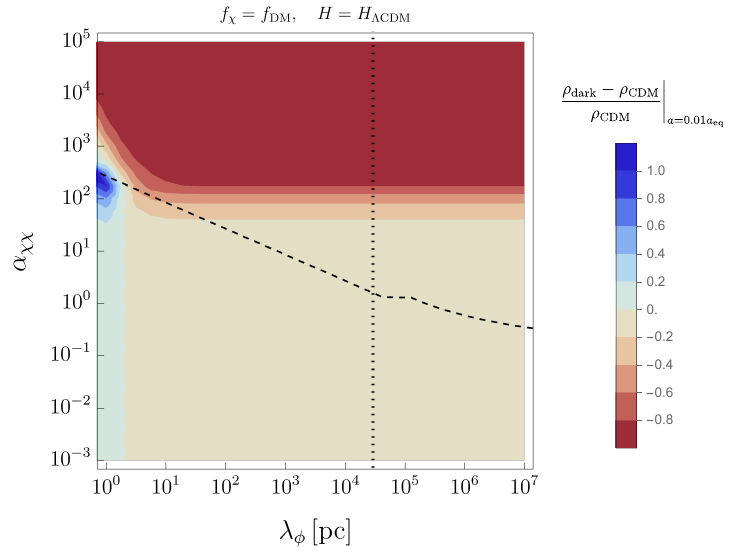}
    \includegraphics[width=0.495\linewidth]{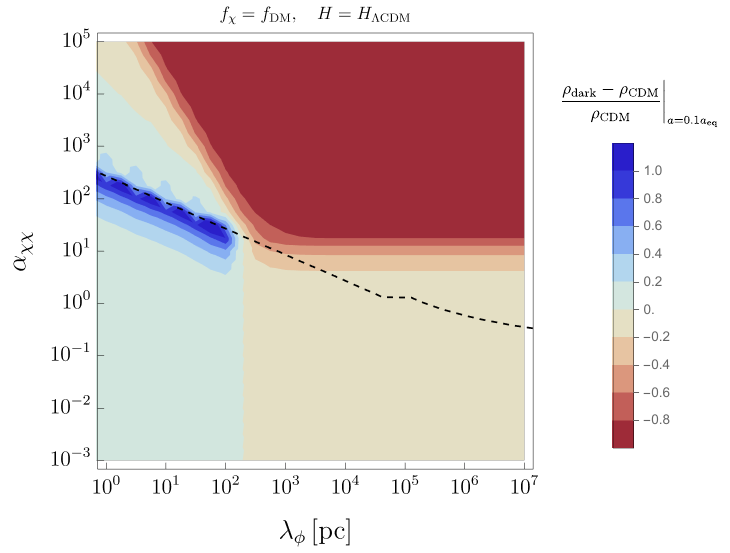}
    \includegraphics[width=0.495\linewidth]{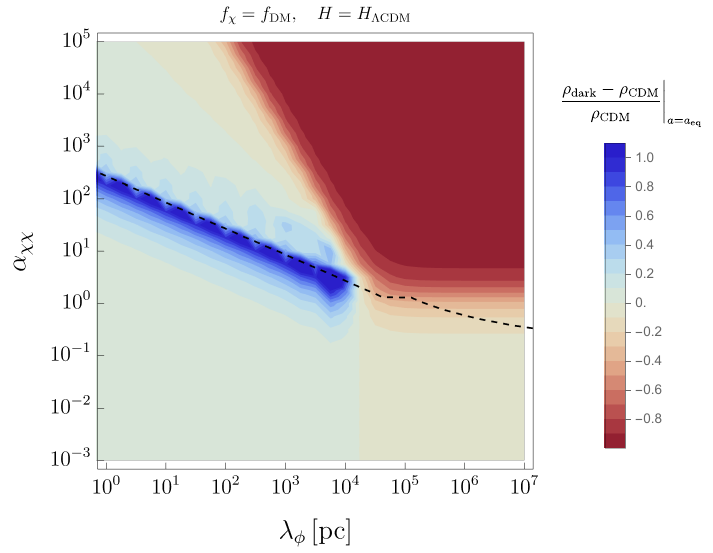}
    \includegraphics[width=0.495\linewidth]{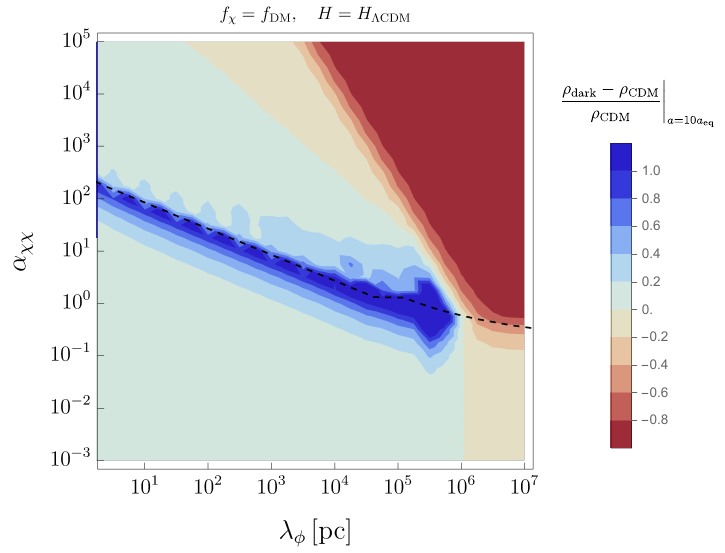}
    \caption{Fractional of $\rho_{\rm dark}$ from $\rho_{\rm CDM}$. 
    The blue region signifies the mixed DM $f_\phi/f_\chi=\mathcal{O}(1)$. The blobby part near the case $\phi_{\rm rise} \phi_0 \phi_* \phi_{\rm DM}$-case $\phi_{\rm rise} \phi_{\rm DM}$ boundary is because the amplitude of $\phi_{\rm DM}$ is set by the relative field-space velocity $\dot{\phi}_{\rm lin-osc}-\dot{\phi}_*$ when $\phi_*\approx\phi_0$, and this relative velocity depends on the phase of $\phi_{\rm lin-osc}$ at that time.  In the bottom right figure, $\lambda_{\phi}\lesssim 1\text{ pc}$ was not simulated because that regime is computationally expensive.}
    \label{fig:rhodarkminrhoCDM}
\end{figure*}

We compare the conservative limits on $\bar{w}_{\rm dark}^{(i)}$ assumed in our analysis (Eq.~\eqref{eq:wlimits}) with the allowed values of $\bar{w}_{\rm dark}^{(i)}$ found by Ref.~\cite{Ilic:2020onu} in Fig.~\ref{fig:wlimitcomparison}. We show the background limits (discussed in subsection \ref{section:EoSconstraints} and subsection.~\ref{ss:backgroundresults}) from each of the four bins in Fig.~\ref{fig:wlimitsperbin}. In Fig.~\ref{fig:rhodarkminrhoCDM}, we show contours of the fractional difference between the dark sector's energy density $\rho_{\rm dark}$ and the CDM value $\rho_{\rm CDM}$ on the $(\lambda_\phi,\alpha_{\chi\chi})$ parameter space, at scale factors $0.01a_{\rm eq},0.1 a_{\rm eq},a_{\rm eq}, 10 a_{\rm eq}$. These contour plots give a rough indication for which parts of the parameter space have the dark sector deviating considerably from CDM at various epochs.

\clearpage


\begin{thebibliography}{10}

\bibitem{Rogers:2020ltq}
K.~K. Rogers and H.~V. Peiris,
\newblock Phys. Rev. Lett. {\bf 126}, 071302 (2021), 2007.12705.

\bibitem{Das:2020nwc}
S.~Das and E.~O. Nadler,
\newblock Phys. Rev. D {\bf 103}, 043517 (2021), 2010.01137.

\bibitem{poulin2016fresh}
V.~Poulin, P.~D. Serpico, and J.~Lesgourgues,
\newblock Journal of Cosmology and Astroparticle Physics {\bf 2016}, 036 (2016).

\bibitem{Graham:2024hah}
P.~W. Graham and H.~Ramani,
\newblock Phys. Rev. D {\bf 110}, 075012 (2024), 2404.01378.

\bibitem{Chang:2024fol}
J.~H. Chang, P.~J. Fox, and H.~Xiao,
\newblock JCAP {\bf 08}, 023 (2024), 2406.09499.

\bibitem{Tulin:2017ara}
S.~Tulin and H.-B. Yu,
\newblock Phys. Rept. {\bf 730}, 1 (2018), 1705.02358.

\bibitem{Slone:2021nqd}
O.~Slone, F.~Jiang, M.~Lisanti, and M.~Kaplinghat,
\newblock Phys. Rev. D {\bf 107}, 043014 (2023), 2108.03243.

\bibitem{Sundrum:2003yt}
R.~Sundrum,
\newblock (2003), hep-th/0312212.

\bibitem{damour1994string}
T.~Damour and A.~M. Polyakov,
\newblock General Relativity and Gravitation {\bf 26}, 1171 (1994).

\bibitem{damour1994string1}
T.~Damour and A.~M. Polyakov,
\newblock The string dilation and a least coupling principle, 1994.

\bibitem{Bogorad:2023wzn}
Z.~Bogorad, P.~W. Graham, and H.~Ramani,
\newblock JCAP {\bf 03}, 067 (2025), 2311.07648.

\bibitem{Keselman:2009oaz}
J.~A. Keselman, A.~Nusser, and P.~J.~E. Peebles,
\newblock Phys. Rev. D {\bf 80}, 063517 (2009), 0902.3452.

\bibitem{Kesden:2006zb}
M.~Kesden and M.~Kamionkowski,
\newblock Phys. Rev. Lett. {\bf 97}, 131303 (2006), astro-ph/0606566.

\bibitem{Kesden:2006vz}
M.~Kesden and M.~Kamionkowski,
\newblock Phys. Rev. D {\bf 74}, 083007 (2006), astro-ph/0608095.

\bibitem{Frieman:1991zxc}
J.~A. Frieman and B.-A. Gradwohl,
\newblock Phys. Rev. Lett. {\bf 67}, 2926 (1991).

\bibitem{1992ApJ...398..407G}
B.-A. {Gradwohl} and J.~A. {Frieman},
\newblock \apj {\bf 398}, 407 (1992).

\bibitem{Farrar:2003uw}
G.~R. Farrar and P.~J.~E. Peebles,
\newblock Astrophys. J. {\bf 604}, 1 (2004), astro-ph/0307316.

\bibitem{Bean:2008ac}
R.~Bean, E.~E. Flanagan, I.~Laszlo, and M.~Trodden,
\newblock Phys. Rev. D {\bf 78}, 123514 (2008), 0808.1105.

\bibitem{Bean:2007ny}
R.~Bean, E.~E. Flanagan, and M.~Trodden,
\newblock Phys. Rev. D {\bf 78}, 023009 (2008), 0709.1128.

\bibitem{2010PhRvD..81f3521K}
J.~A. {Keselman}, A.~{Nusser}, and P.~J.~E. {Peebles},
\newblock \prd {\bf 81}, 063521 (2010), 0912.4177.

\bibitem{Hellwing:2008qf}
W.~A. Hellwing and R.~Juszkiewicz,
\newblock Phys. Rev. D {\bf 80}, 083522 (2009), 0809.1976.

\bibitem{Bottaro:2024pcb}
S.~Bottaro, E.~Castorina, M.~Costa, D.~Redigolo, and E.~Salvioni,
\newblock Phys. Rev. D {\bf 112}, 023525 (2025), 2407.18252.

\bibitem{Bottaro:2023wkd}
S.~Bottaro, E.~Castorina, M.~Costa, D.~Redigolo, and E.~Salvioni,
\newblock Phys. Rev. Lett. {\bf 132}, 201002 (2024), 2309.11496.

\bibitem{Archidiacono:2022iuu}
M.~Archidiacono, E.~Castorina, D.~Redigolo, and E.~Salvioni,
\newblock JCAP {\bf 10}, 074 (2022), 2204.08484.

\bibitem{Costa:2025kwt}
M.~Costa, C.~Creque-Sarbinowski, O.~Simon, and Z.~J. Weiner,
\newblock (2025), 2510.00098.

\bibitem{Domenech:2021uyx}
G.~Dom{\`e}nech and M.~Sasaki,
\newblock JCAP {\bf 06}, 030 (2021), 2104.05271.

\bibitem{Domenech:2023afs}
G.~Dom{\`e}nech, D.~Inman, A.~Kusenko, and M.~Sasaki,
\newblock Phys. Rev. D {\bf 108}, 103543 (2023), 2304.13053.

\bibitem{Afshordi:2005ym}
N.~Afshordi, M.~Zaldarriaga, and K.~Kohri,
\newblock Phys. Rev. D {\bf 72}, 065024 (2005), astro-ph/0506663.

\bibitem{Savastano:2019zpr}
S.~Savastano, L.~Amendola, J.~Rubio, and C.~Wetterich,
\newblock Phys. Rev. D {\bf 100}, 083518 (2019), 1906.05300.

\bibitem{Blinov:2016kte}
N.~Blinov and A.~Hook,
\newblock JHEP {\bf 06}, 176 (2016), 1605.03178.

\bibitem{Hook:2018jle}
A.~Hook,
\newblock Phys. Rev. Lett. {\bf 120}, 261802 (2018), 1802.10093.

\bibitem{Brzeminski:2020uhm}
D.~Brzeminski, Z.~Chacko, A.~Dev, and A.~Hook,
\newblock Phys. Rev. D {\bf 104}, 075019 (2021), 2012.02787.

\bibitem{Planck:2018vyg}
Planck, N.~Aghanim {\em et~al.},
\newblock Astron. Astrophys. {\bf 641}, A6 (2020), 1807.06209,
\newblock [Erratum: Astron.Astrophys. 652, C4 (2021)].

\bibitem{Hu:2000ke}
W.~Hu, R.~Barkana, and A.~Gruzinov,
\newblock Phys. Rev. Lett. {\bf 85}, 1158 (2000), astro-ph/0003365.

\bibitem{Ferreira:2020fam}
E.~G.~M. Ferreira,
\newblock Astron. Astrophys. Rev. {\bf 29}, 7 (2021), 2005.03254.

\bibitem{Hui:2021tkt}
L.~Hui,
\newblock Ann. Rev. Astron. Astrophys. {\bf 59}, 247 (2021), 2101.11735.

\bibitem{Ilic:2020onu}
S.~Ili{\'c}, M.~Kopp, C.~Skordis, and D.~B. Thomas,
\newblock Phys. Rev. D {\bf 104}, 043520 (2021), 2004.09572.

\bibitem{Kopp:2018zxp}
M.~Kopp, C.~Skordis, D.~B. Thomas, and S.~Ili{\'c},
\newblock Phys. Rev. Lett. {\bf 120}, 221102 (2018), 1802.09541.

\bibitem{2012PhRvD..86l3504S}
J.~{Samsing}, E.~V. {Linder}, and T.~L. {Smith},
\newblock \prd {\bf 86}, 123504 (2012), 1208.4845.

\bibitem{Hojjati:2013oya}
A.~Hojjati, E.~V. Linder, and J.~Samsing,
\newblock Phys. Rev. Lett. {\bf 111}, 041301 (2013), 1304.3724.

\bibitem{Hu:1998kj}
W.~Hu,
\newblock Astrophys. J. {\bf 506}, 485 (1998), astro-ph/9801234.

\bibitem{1974A&A....37..225M}
P.~{Meszaros},
\newblock \aap {\bf 37}, 225 (1974).

\bibitem{Hu:1995en}
W.~Hu and N.~Sugiyama,
\newblock Astrophys. J. {\bf 471}, 542 (1996), astro-ph/9510117.

\bibitem{Graham:2023unf}
P.~W. Graham and H.~Ramani,
\newblock Phys. Rev. D {\bf 110}, 075011 (2024), 2311.07654.

\bibitem{Graham:2025fdt}
P.~W. Graham, D.~Green, and J.~Meyers,
\newblock (2025), 2508.20999.

\bibitem{Iovino:2024tyg}
A.~J. Iovino, G.~Perna, A.~Riotto, and H.~Veerm{\"a}e,
\newblock JCAP {\bf 10}, 050 (2024), 2406.20089.

\bibitem{Bringmann:2025cht}
T.~Bringmann, D.~Croon, and S.~Sevillano~Mu{\~n}oz,
\newblock (2025), 2506.20704.

\bibitem{Efstathiou:1998xx}
G.~Efstathiou and J.~R. Bond,
\newblock Mon. Not. Roy. Astron. Soc. {\bf 304}, 75 (1999), astro-ph/9807103.

\bibitem{Chakraborty:2022mwu}
A.~Chakraborty, P.~K. Chanda, K.~L. Pandey, and S.~Das,
\newblock Astrophys. J. {\bf 932}, 119 (2022), 2204.09628.

\bibitem{Chakraborty:2024xas}
A.~Chakraborty, T.~Ray, S.~Das, A.~Banerjee, and V.~Ganesan,
\newblock (2024), 2403.14247.

\bibitem{Fardon:2003eh}
R.~Fardon, A.~E. Nelson, and N.~Weiner,
\newblock JCAP {\bf 10}, 005 (2004), astro-ph/0309800.

\bibitem{Kaplan:2004dq}
D.~B. Kaplan, A.~E. Nelson, and N.~Weiner,
\newblock Phys. Rev. Lett. {\bf 93}, 091801 (2004), hep-ph/0401099.

\bibitem{Sakstein:2019fmf}
J.~Sakstein and M.~Trodden,
\newblock Phys. Rev. Lett. {\bf 124}, 161301 (2020), 1911.11760.

\bibitem{Berryman:2022hds}
J.~M. Berryman {\em et~al.},
\newblock Phys. Dark Univ. {\bf 42}, 101267 (2023), 2203.01955.

\bibitem{Banerjee:2025nvs}
A.~Banerjee, N.~H. Nguyen, and E.~H. Tanin,
\newblock (2025), 2509.25308.

\bibitem{Bogorad:2024hfj}
Z.~Bogorad, P.~W. Graham, and H.~Ramani,
\newblock JCAP {\bf 04}, 006 (2025), 2410.07324.

\bibitem{Poulin:2018cxd}
V.~Poulin, T.~L. Smith, T.~Karwal, and M.~Kamionkowski,
\newblock Phys. Rev. Lett. {\bf 122}, 221301 (2019), 1811.04083.

\bibitem{Kamionkowski:2022pkx}
M.~Kamionkowski and A.~G. Riess,
\newblock Ann. Rev. Nucl. Part. Sci. {\bf 73}, 153 (2023), 2211.04492.

\bibitem{Inomata:2018epa}
K.~Inomata and T.~Nakama,
\newblock Phys. Rev. D {\bf 99}, 043511 (2019), 1812.00674.

\bibitem{Ma:1994dv}
C.-P. Ma and E.~Bertschinger,
\newblock (1994), astro-ph/9401007.

\bibitem{Tanin:2017bzm}
E.~H. Tanin and E.~D. Stewart,
\newblock JCAP {\bf 11}, 019 (2017), 1708.04865.

\bibitem{Banerjee:2025qyg}
A.~Banerjee, N.~H. Nguyen, and E.~H. Tanin,
\newblock (2025), 2512.05188.

\end{thebibliography}

\end{document}